\documentclass[manuscript]{aa}
\usepackage{graphicx}
\usepackage[varg]{txfonts}
\usepackage{hyperref}

\begin{document}

\title{Modeling the thermal conduction in the solar atmosphere with the code MANCHA3D}

\titlerunning{Modeling the thermal conduction with the code MANCHA3D}
\authorrunning{A. Navarro et al.}

\author{Anamar\'ia Navarro\and 
E. Khomenko\and 
M. Modestov\and 
N. Vitas 
}

\institute{Instituto de Astrof\'isica de Canarias, 38205 La Laguna, Tenerife, Spain 
\and
Dpto. de Astrof\'isica, Universidad de La Laguna, 38205 La Laguna, Tenerife, Spain \email{anavarro@iac.es} 
}



\abstract{
Thermal conductivity is one of the important mechanisms of heat transfer in the solar corona. In the limit of strongly magnetized plasma, it is typically modeled by Spitzer's expression where the heat flux is aligned with the magnetic field.}
{
This paper describes the implementation of the heat conduction into the code MANCHA3D with an aim of extending single-fluid MHD simulations from the upper convection zone into the solar corona.}
{
Two different schemes to model heat conduction are implemented: (1) a standard scheme where a parabolic term is added to the energy equation, and (2) a scheme where the hyperbolic heat flux equation is solved.}
{
The first scheme limits the time step due to the explicit integration of a parabolic term, which makes the simulations computationally expensive. The second scheme solves the limitations on the time step by artificially limiting the heat conduction speed to computationally manageable values. The validation of both schemes is carried out with standard tests in one, two, and three spatial dimensions. Furthermore, we implement the model for heat flux derived by \citet{1965RvPP....1..205B} in its most general form, when the expression for the heat flux depends on the ratio of the collisional to cyclotron frequencies of the plasma, and, therefore on the magnetic field strength. Additionally, our implementation takes into account the heat conduction in parallel, perpendicular, and transverse directions, and provides the contributions from ions and electrons separately. The model recovers Spitzer's expression for parallel thermal conductivity in the strongly magnetized limit but also transitions smoothly between field-aligned conductivity and isotropic conductivity for regions with a low or null magnetic field.
We describe the details of the implementation of Braginskii's thermal conductivity using a combination of the first scheme for the perpendicular and transverse directions and the second scheme for the parallel component. We estimate thermal conductivities in a quiet-Sun model. In this model, we find that the perpendicular and transverse components for electrons and ions and the parallel component for ions might have some significance below the transition region. Above the transition region only the parallel component for ions might be important. Finally, we present a two-dimensional test for heat conduction using realistic values of the solar atmosphere where we prove the robustness of the two schemes implemented and show that our adaptation of the hyperbolic treatment offers a great advantage over the computational cost of the simulations.}
{
}

\maketitle

\section{Introduction}\label{sec:intro}

Thermal conductivity is an essential physical process by which pressure and temperature perturbations are spread out through a plasma. It corresponds to a macroscopic manifestation of a transport process (energy transfer by particle collisions) and its strength depends on the temperature and temperature gradient.  It plays an important role in diverse range of conditions from those found in  thermoelectric materials \citep{2022FrPhy..1723202Y, 2022JPCS..16110390L} to supernova remnants \citep{10.1111/j.1365-2966.2008.13085.x}. In the solar atmosphere, thermal conduction is one of the dominant processes that shapes the dynamics of the hot plasma \citep{2020A&C....3000341Y}. For instance, it contributes to the energy redistributions in plasma eruptions \citep{2012ApJ...758...53B, Liu_2009, 2020A&C....3000341Y, 2021MNRAS.500.3329N}. Heat conduction is a key aspect in the numerical analysis of the coronal heating problem \citep{ 2011A&A...530A.112B,2011A&A...531A.154G, 2013A&A...550A..30B,2013A&A...555A.123B, 2014A&A...564A..12C}.

The ratio of the collisional frequency to the cyclotron frequency, which depends on the magnetic field, determines the direction of the heat flux with respect to the magnetic field direction. In the corona, where the magnetic field is strong and the collision frequency is low, the thermal conduction is aligned with the magnetic field and is commonly modeled by Spitzer's expression for fully ionized plasmas. Likewise, \citet{1965RvPP....1..205B} developed a more complete description of the heat flux which depends on the ratio between the cyclotron frequency and the collisional frequency of the plasma. In his model, the heat flux is  decomposed into three parts, a parallel component, a perpendicular component and a transverse component, each of them for ions and electrons. Therefore it allows a smooth transition between field-aligned conductivity and isotropic conductivity in regions with a low or null magnetic field or regions with high collisional frequency. Furthermore, the parallel component of the electron heat flux has roughly the same value as the Spitzer's expression. Accordingly, \citet{1965RvPP....1..205B} model is appropriate for the description of the complete solar atmosphere.

Modeling the thermal conductivity numerically is a challenging endeavor since explicit solvers for a parabolic problem lead to very small time-steps according to the Courant–Friedrichs–Lewy (CFL) condition in comparison to a hyperbolic problem defined by ideal magnetohydrodynamics (MHD) equations. Different schemes have been developed to overcome this restriction, for instance: Super Time-Stepping methods \citep{10.1111/j.1365-2966.2012.20744.x} \citep[used in the MPI-AMRVAC code, see][]{2018ApJS..234...30X}, implicit–explicit schemes \citep{10.1111/j.1365-2966.2008.13085.x},  semi-implicit directionally-split methods \citep{2011JCoPh.230.4899S, 2020A&C....3000341Y}, symmetric and asymmetric discretization schemes \citep{2005JCoPh.209..354G, 2005ApJ...633..334P}. With the same purpose, \cite{2017ApJ...834...10R} and \cite{2020GApFD.114..261W} introduced an efficient way to deal with the numerical time step constraints by solving the hyperbolic equation for the heat flux for the MURAM and PENCIL codes respectively. In the current work, we implemented into the code  MANCHA3D \citep{2010ApJ...719..357F, 2018A&A...615A..67G, 2018A&A...618A..87K} a combination of both schemes, the hyperbolic equation to evolve the parallel to the magnetic field component of the heat flux, and the parabolic equation for the perpendicular and transverse directions.

The paper is organized as follows. In Section \ref{sec:equations} we describe two different schemes for the implementation of thermal conductivity in the code. From Sections \ref{sec:test1d} to \ref{sec:ring3d} we validate the numerical schemes with different tests in one, two and three dimensions. The general Braginskii heat flux is presented in Section \ref{sec:braginskii}, we describe its implementation in the code and compare the thermal conductivity in parallel, perpendicular and transverse directions, from ions and electrons in a quiet Sun stratified atmosphere. Finally, in Section \ref{sec:hottop} we propose a two-dimensional test with a realistic solar stratified atmosphere and in Section \ref{sec:conclusions} we present our conclusions.


\section{Brief description of the MANCHA3D code} \label{sec:mancha3D}

MANCHA3D  is a three-dimensional numerical code designed to simulate complex dynamics of the solar atmosphere plasma. It solves the non-linear MHD equations taking into account non ideal effects as the ambipolar diffusion, Hall effect, the battery effect and radiative losses, although, in the simulations that are described in this work, such effects are not taken into account. A gravitationally stratified plasma in the ideal regime can be represented by the following set of equations
\begin{eqnarray}
& & \frac{\partial \rho}{\partial t} + \nabla \cdot \left( \rho {\bf v} \right) = 0 \, , \label{eq:rho} \\
& & \frac{\partial{\bf B}}{\partial t} = \nabla \times \left( {\bf v} \times {\bf B} \right) \, , \label{eq:B}  \\
& & \frac{\partial (\rho {\bf v})}{\partial t} + \nabla \cdot \left[   \rho {\bf v} {\bf v} + \left( p + \frac{{\bf B}^2}{2\mu_0} \right) {\bf I} - \frac{{\bf B B} }{\mu_0}  \right] = \rho {\bf g} \, ,  \label{eq:rhoV}  \\
& & \frac{\partial e}{\partial t} + \nabla \cdot \left[   \left(  e + p + \frac{{\bf B}^2}{2 \mu_0} \right){\bf v}   - \frac{\bf B}{\mu_0} \left(  {\bf B \cdot v} \right)   \right]  = \rho {\bf v} \cdot {\bf g}   \, , \label{eq:energy} 
\end{eqnarray}
where $\rho$ is the density, ${\bf v}$ is the velocity, $p$ is the gas pressure, ${\bf B}$ is the magnetic field, ${\bf g}$ is the gravitational acceleration, $e$ is the total energy per unit volume and ${\bf I}$ is the identity tensor The dot ``$\cdot$'' represents the scalar product of vectors, while the notation ``{\bf B}{\bf B}'' stands for the tensor product. 

In the code, the equations are discretized with a centered finite differences scheme of sixth order and integrated in time with an explicit Runge–Kutta method. For numerical stability, the code uses hyperdiffusivity and filtering. A more complete description of the code can be found in the works of \cite{2006ApJ...653..739K, 2008ApJ...689.1379K, 2010ApJ...719..357F, 2018A&A...615A..67G, 2018A&A...618A..87K}.

\section{Numerical schemes for thermal conductivity} \label{sec:equations}

The effects of thermal conductivity can be represented by adding the divergence of the heat flux vector {\bf q} into the energy equation (\ref{eq:energy}). Accordingly, the new equation for the energy reads as
\begin{eqnarray}
& & \frac{\partial e}{\partial t} + \nabla \cdot \left[   \left(  e + p + \frac{{\bf B}^2}{2 \mu_0} \right){\bf v}   - \frac{\bf B}{\mu_0} \left(  {\bf B \cdot v} \right)   \right] = \rho {\bf v} \cdot {\bf g} - \nabla \cdot {\bf q}  \, . \label{eq:E_q}
\end{eqnarray}

In the strongly magnetized limit, for the fully ionized plasma, the heat flux is aligned with the magnetic field
\begin{eqnarray}
{\bf q} = - \kappa_{\|} \nabla_{\|}T =  - \kappa_{0} T^{5/2} {\bf \hat{b}} \left( {\bf \hat{b}} \cdot {\bf \nabla }  \right) T   \, , \label{eq:q_spitzer}
\end{eqnarray}
where ${\bf \hat{b}}$ is the unit vector along the magnetic field and the term $({\bf \hat{b}} \cdot \nabla T)$ represents the temperature gradient along the magnetic field, $\kappa_{||}$ is the parallel conductivity and $\kappa_{0}$ is a constant that depends on the properties of the plasma. This equation is referred as Spitzer's thermal conduction vector \citep{Spitzer1956}. 

The integration time step of a fully explicit finite difference method is limited by the CFL condition. For the heat conduction problem, the value of the limiting time step can be obtained by comparing the time-derivative and the thermal conduction terms in the energy equation, assuming no variations in density and the ideal gas law 
\begin{eqnarray}
\frac{\partial e}{\partial t} \approx \frac{(\gamma-1) \kappa_{0} T^{7 / 2}}{p} \nabla^{2} e \, ,
\end{eqnarray}
therefore, the time step is given by
\begin{eqnarray}
\mathrm{d}t \leq \text{cfl}_{\text{TC}} \frac{\mathrm{min}(\Delta x, \Delta y, \Delta z)^2}{\text{max}((\gamma-1) \kappa_{0} T^{7/2}/p)} \, ,
\end{eqnarray}
where $\Delta x$, $\Delta y$ and $\Delta z$ are the grid sizes and $\text{cfl}_{\text{TC}}$ is a stability constant that takes values of 0.5, 0.5 and 1/3 for 1D, 2D and 3D cases, respectively. 

In some cases, the thermal conduction time step
can become much smaller than the time step given by the MHD CFL condition. To overcome this difficulty, \cite{2017ApJ...834...10R} and \cite{2020GApFD.114..261W} proposed to solve the hyperbolic equation for the heat flux instead of the classical parabolic equation, i.e., 
\begin{eqnarray} \label{eq:qhyper}
& &\frac{\partial q}{\partial t} = \frac{1}{\tau} \left(  - f_{\text{sat}} \kappa_{0} T^{5/2} \left( {\bf \hat{b}} \cdot \nabla \right) T - q \right) \, , \label{eq:q_evolution} \\
& &  \frac{\partial e}{\partial t} = \left[ \dots \right] - \nabla \cdot \left( q \hat{\bf b} \right) \, .  
\end{eqnarray}
The factor $f _{\text{sat}}$ sets the saturation of the conductive heat flux, preventing it from taking highest values, and  $\tau$ is  the relaxation time. Following \cite{1985ApJ...289..414F} and \cite{10.1111/j.1365-2966.2012.20744.x} the saturation factor can be written as
\begin{eqnarray}
f_{\text{sat}} = \left(  1 + \frac{|\kappa_{0} T^{5/2} \left( \hat{\bf b} \cdot \nabla \right) T|}{1.5 \rho c_s^3}  \right)^{-1} \, , \label{eq:fsat}
\end{eqnarray}
where $c_s = \sqrt{\gamma p / \rho }$ denotes the speed of sound and $\gamma$ is the adiabatic index.

The hyperbolic heat-conduction equation corresponds to the modification of the classical Fourier heat flux by the addition of an additional extra thermal inertia term. This modification was proposed independently by \cite{vernotte1958paradoxes} and \cite{cattaneo1958forme} in order to suppress the paradox of the infinite speed of propagation of the heat conduction in the Fourier theory, and implying the existence of thermal waves. This approach has been widely studied, both theoretically and experimentally,  see \cite{ABDELHAMID1999899} and references therein.

In Eq. \ref{eq:q_evolution} the relaxation time depends on the mechanism of heat transport, and represents the time lag needed to establish steady-state heat conduction in an element of volume when a temperature gradient is suddenly applied to that element \citep{ABDELHAMID1999899}. Different definitions for the relaxation time are proposed by \cite{2017ApJ...834...10R} and \cite{2020GApFD.114..261W}, however they do not allow stable running of all our samples, therefore we use a fixed value for the relaxation time, $\tau = 4 \mathrm{d}t$ in Eq. \ref{eq:qhyper}.

\section{Tests for validation of the numerical scheme}

\subsection{One-dimensional problem} \label{sec:test1d}

For the first test, we have considered the following one-dimensional heat conduction problem proposed by \cite{2017ApJ...834...10R}. The initial non-equilibrium temperature profile is given by
\begin{eqnarray}
T_{0}(x) = 0.1 + 0.9 x^{5} \, ,
\end{eqnarray}
in non-dimensional units. In these units, we use density $\rho = 1.0$, magnetic field $B_x = 1.0$, $B_y = Bz = 0.0$, and thermal conductivity coefficient $\kappa_0 = 1.0$. The domain extends over $0 \leq x \leq 1$ and is covered by 250 points. The boundary conditions set all the quantities to be fixed in time. The asymptotic stationary solution is given by
\begin{eqnarray}
T(x)=\left[0.1^{3.5}+\left(1-0.1^{3.5}\right) x\right]^{2 / 7} \, . \label{eq:asymptotic_solution}
\end{eqnarray}

To analyze this problem, we perform four different simulations. On the one hand, we have performed a run with a standard parabolic scheme (we will refer to this scheme as scheme \#1 over the paper), where we solve the equations (\ref{eq:rho}), (\ref{eq:B}), (\ref{eq:rhoV}) and (\ref{eq:E_q}) using the field-aligned heat flux
\begin{eqnarray}
{\bf q} = - f_{\text{sat}} \kappa_{0} T^{5/2} {\bf \hat{b}} \left( {\bf \hat{b}} \cdot {\bf \nabla }  \right) T  \, \label{eq:q_sat} . 
\end{eqnarray}
Here we have included the saturation factor $f_{\text{sat}}$ given by (\ref{eq:fsat}) to the Spitzer heat flux (\ref{eq:q_spitzer}). This is needed for consistency in comparison of the simulations with different schemes. On the other hand, we run simulations with the hyperbolic scheme using three different time steps, $\mathrm{d}t = \mathrm{d}t_{\text{TC}}$, $\mathrm{d}t = 50 \mathrm{d}t_{\text{TC}}$ and $\mathrm{d}t = 100 \mathrm{d}t_{\text{TC}}$. We will refer to the latter one as scheme \#2. 

\begin{figure*}
 \centering
 \includegraphics[width = 0.48\textwidth]{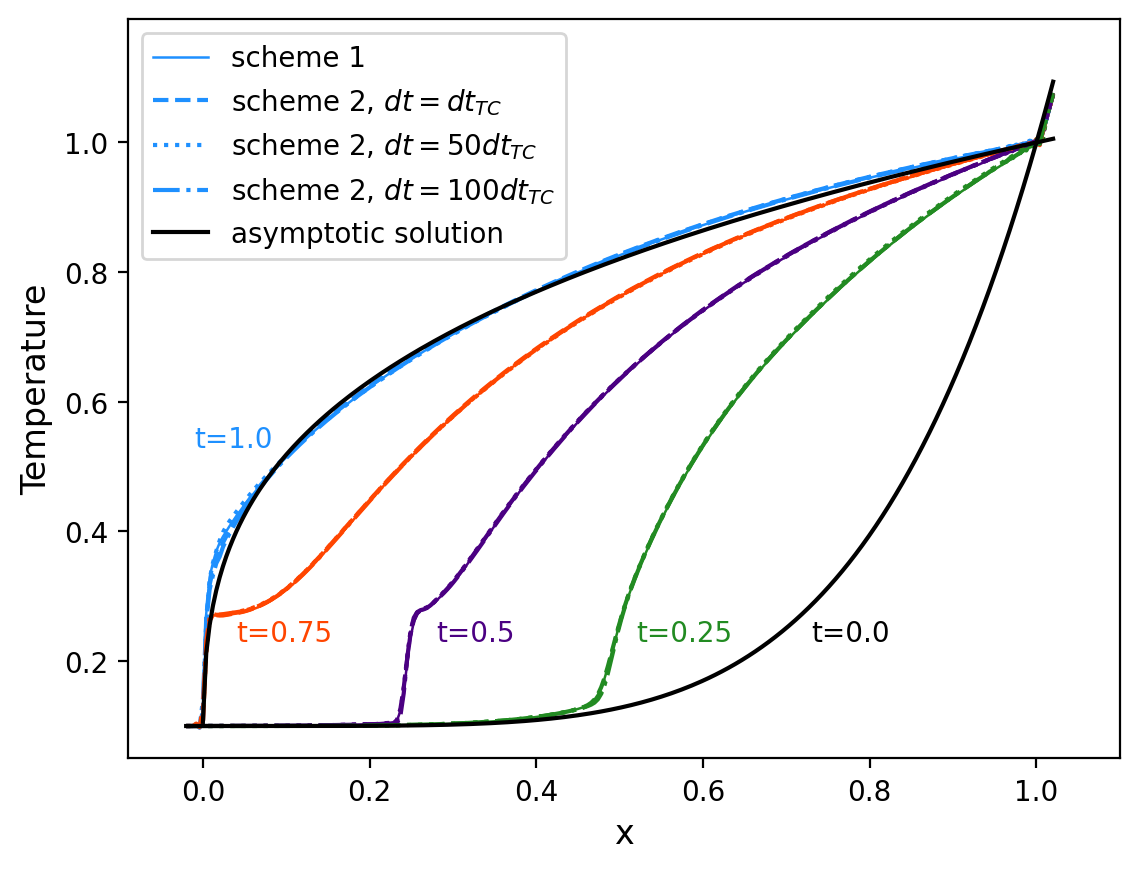}
  \includegraphics[width = 0.48\textwidth]{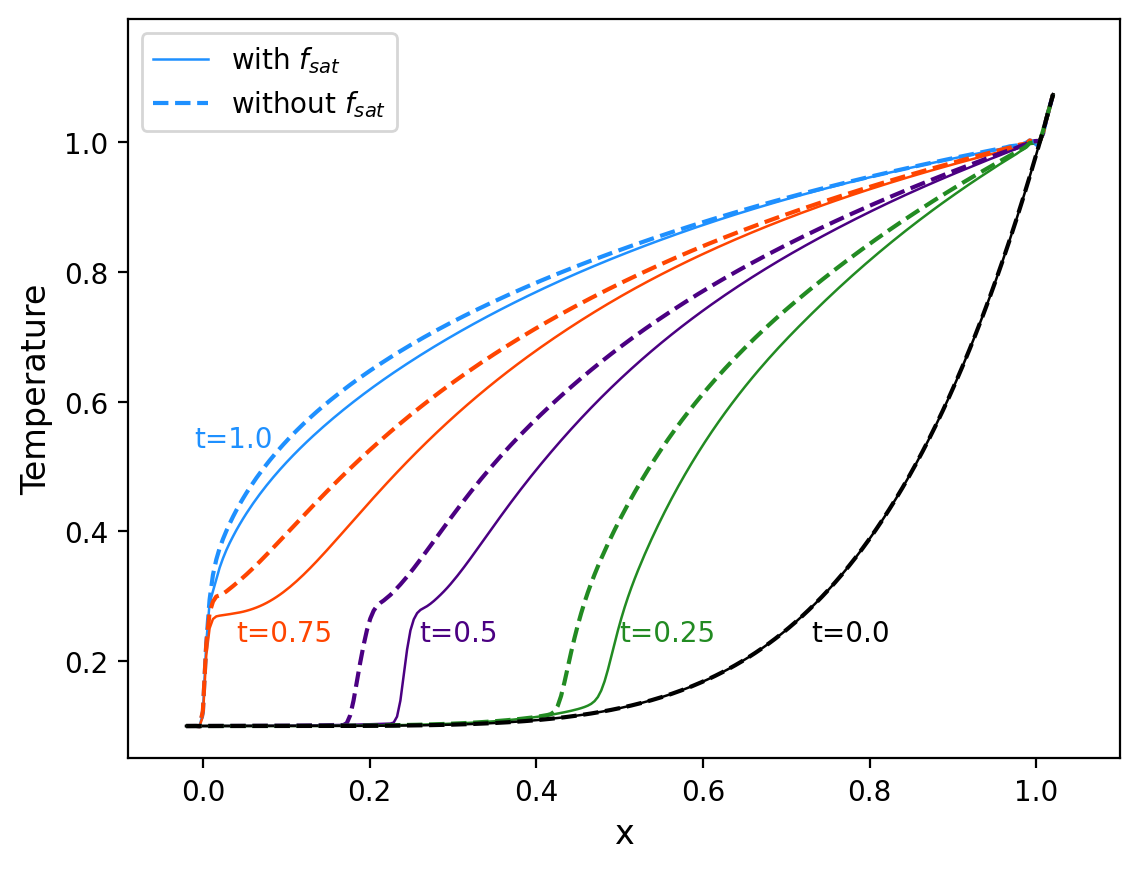}
 \caption{Left panel shows the evolution of temperature in the one-dimensional thermal conductivity test computed with the scheme \#1 (explicit evolution of the parabolic term) and with the scheme \#2 (hyperbolic equation for heat flux) with three different time steps. Right panel shows a comparison between simulations using scheme \# 2, one including the saturation factor of Eq. (\ref{eq:fsat}) and another one without it}
 \label{fig:test_1D}
\end{figure*}

Figure \ref{fig:test_1D} shows different stages of the time evolution, $t=0.25$, $t=0.5$, $t=0.75$ and $t=1.0$  for all the simulations. The solid black lines represent the initial state ($t=0$) and the asymptotic solution (\ref{eq:asymptotic_solution}). In all the cases, the reference solution is recovered around $t=1$ and the intermediate steps of the evolution of the temperature are quite similar between different simulation cases. 
It should be emphasized that the temperature evolves with the same speed for both parabolic and hyperbolic schemes, because both schemes use the same saturation factor. 
It decreases the magnitude of the heat flux enhancing stability of the code. 
Without the saturation factor, the temperature evolves faster with respect to the simulation time, but it requires a smaller time step for a stable run. This faster evolution is clearly shown at left panel of Figure \ref{fig:test_1D}. 
The saturation factor allows running simulation with much larger time step, however if the time step is too high, the simulation becomes unstable, as shown in the top left panel of Figure \ref{fig:nonworking}. 

To estimate the accuracy of our results, we compare them to the reference solution (\ref{eq:asymptotic_solution}). Table \ref{tabla_test1d} presents the errors $L^1$, $L^2$ and $L^\infty$, computed according to the following definitions,
\begin{eqnarray}
L_N^1 &=& \frac{1}{N} \sum_{i=1, N} | T_i - T_i^{\text{ref}}| \, , \\ 
L_N^2 &=& \frac{1}{N} \sqrt{ \sum_{i=1, N} | T_i - T_i^{\text{ref}}|^2} \, , \\
L_N^{\infty} &=& \textrm{max}(| T_i - T_i^{\text{ref}}| ) \, , 
\end{eqnarray}
where $N$ is the total number of points in the domain. We computed the values of $L^1$, $L^2$ and $L^\infty$ at $t=1$. The Table gives values for the simulations with the different schemes and, for the scheme \#2, the values for large time steps are presented. The errors slightly increase between scheme \#1 and \#2, and also increase for larger time steps. However, they stay withing the same order of magnitude. It suggests that we can speed up the simulations using the hyperbolic treatment and still get rather accurate modeling of thermal conduction problems. 

\begin{table} 
\centering
\begin{tabular}{|lcccc|}
 \hline 
 Method & $\mathrm{d}t$ & $L^1$  & $L^2$ & $L^{\infty}$  \\ 
\hline
scheme \#1 & $\mathrm{d}t_{\text{TC}}$     &  7.53$\times 10^{-3}$ & 7.25$\times 10^{-4}$ & 6.86$\times 10^{-2}$  \\    
scheme \#2 & $\mathrm{d}t_{\text{TC}}$     &  9.67$\times 10^{-3}$ & 8.37$\times 10^{-4}$ & 6.86$\times 10^{-2}$  \\    
scheme \#2 & $50 \mathrm{d}t_{\text{TC}}$  &  1.26$\times 10^{-2}$ & 1.03$\times 10^{-3}$ & 7.42$\times 10^{-2}$ \\
scheme \#2 & $100 \mathrm{d}t_{\text{TC}}$ &  1.88$\times 10^{-2}$ & 1.36$\times 10^{-3}$ & 8.58$\times 10^{-2}$ 
\\    
\hline      
\end{tabular}\caption{Errors $L^1$, $L^2$ and $L^\infty $ for the one-dimensional conduction test. }  \label{tabla_test1d}  
\end{table} 


\subsection{Two-dimensional Static Ring} \label{sec:ring2d}

The ring diffusion test proposed by \cite{2005ApJ...633..334P} is a standard test to evaluate the monotonicity properties of the anisotropic conduction. The setup consists of hot patch diffusing in a fixed circular-shaped magnetic field. This test is important to check the numerical schemes since the magnetic field lines align in all possible angles with respect to the Cartesian grid.

\begin{figure*}
 \centering
 \includegraphics[height = 0.22\textheight]{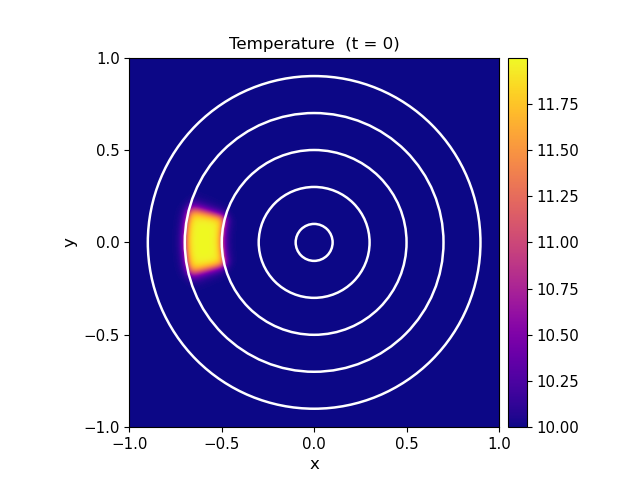}
 \includegraphics[height = 0.22\textheight]{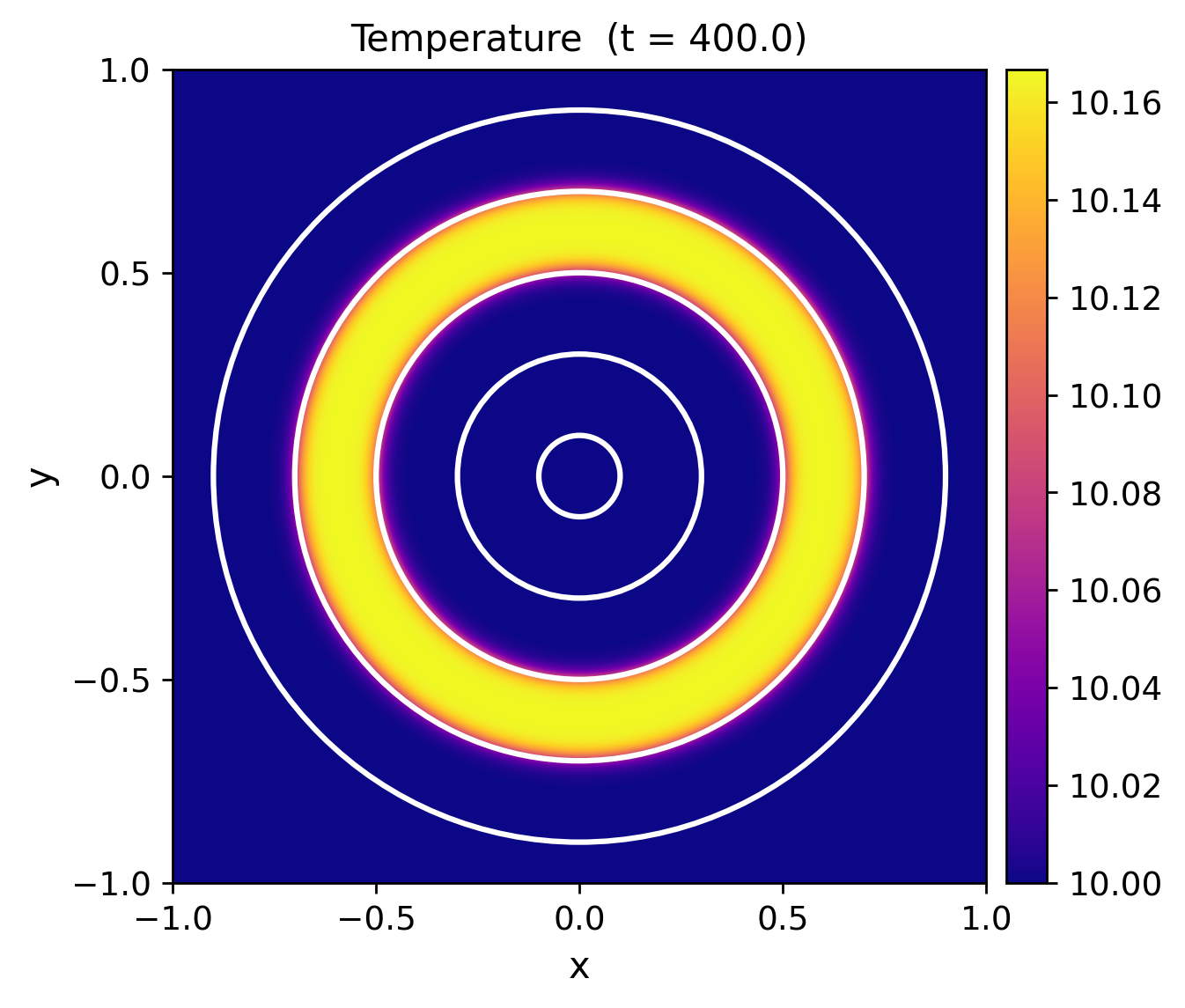} \\
 \includegraphics[height = 0.2\textheight]{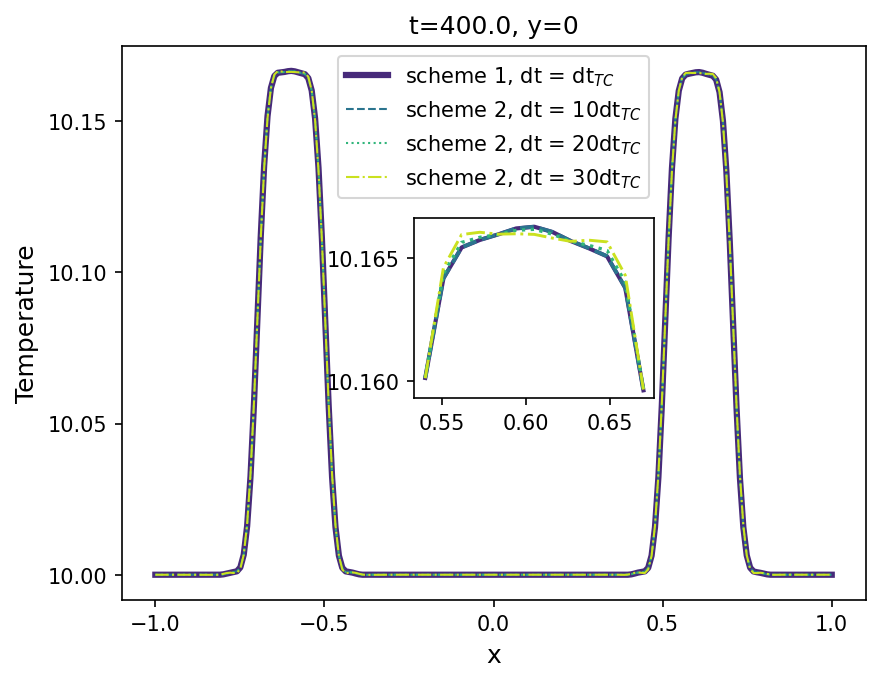}
 \includegraphics[height = 0.2\textheight]{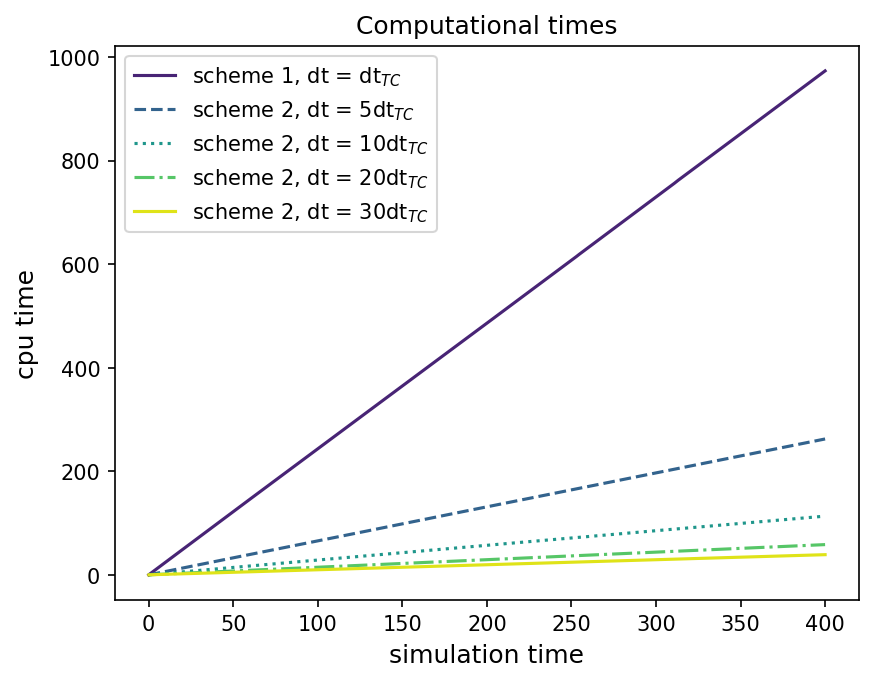} 
 \caption{Two-dimensional static ring problem. Top panels show colormaps of the temperature and magnetic field lines at the initial and final states. The bottom left panel shows a comparison of the temperature at $y = 0$ between the different simulations at the final stage. Bottom right panel gives a graphical representation of the computational times.}
 \label{fig:ring_2D}
\end{figure*}

The initial temperature is given by
\begin{eqnarray}
T = \left\{\begin{array}{ll}
12 & \text { if } 0.5<r<0.7 \text { and } \frac{11}{12} \pi<\theta<\frac{13}{12} \pi \\
10 & \text { otherwise }
\end{array}\right. \, ,
\end{eqnarray}
in non-dimensional units, where $r=\sqrt{x^{2}+y^{2}}$ and $\tan \theta=y / x$. The density is constant and set to $\rho = 1$, and the circular magnetic field is given by
\begin{eqnarray}
B_{x}&=&10^{-5} \cos (\theta+\pi / 2) / r \, , \\
B_{y}&=&10^{-5} \sin (\theta+\pi / 2) / r \, .
\end{eqnarray}
Initially, the system is at rest with $v_{x}=v_{y}=v_{z}=0$.  To study this problem  we use a numerical box of the size $[-1,1] \times[-1,1]$, covered by $200\times 200$ points, we evolve only energy equation with a parallel thermal conductivity to $\kappa_{\|} = 0.01$, and outflow boundary conditions. We run this problem using the scheme \#1 with $\mathrm{d}t = \mathrm{d}t_{\text{TC}}$ and, to evaluate the scheme \#2, we performed simulations using the time steps $\mathrm{d}t =  \mathrm{d}t_{\text{TC}}$ and $\mathrm{d}t = 10 \mathrm{d}t_{\text{TC}}$, $\mathrm{d}t = 20 \mathrm{d}t_{\text{TC}}$, and $\mathrm{d}t = 30 \mathrm{d}t_{\text{TC}}$. Top panels of Figure \ref{fig:ring_2D} show the colormaps of the temperature and the magnetic field lines at the initial time moment $(t=0)$ on the left and at the final stage $(t = 400)$ on the right.

The bottom left plot of Figure \ref{fig:ring_2D} compares the temperature at $y=0$ for the different runs, and shows nearly identical results. The panel at the bottom right of Figure \ref{fig:ring_2D} displays the computational times for the different numerical schemes and it clearly demonstrates that the hyperbolic treatment is much more efficient then the parabolic one. When we set the time factor $\mathrm{d}t = 10 \mathrm{d}t_{\text{TC}}$ the simulation is 10.2 times faster compared with the scheme \#1. Accordingly, when $\mathrm{d}t = 20 \mathrm{d}t_{\text{TC}}$ the efficiency increases 20.3 times and for $\mathrm{d}t = 30 \mathrm{d}t_{\text{TC}}$, 30.3 times.  An unstable example for this test is given at the top right panel of Figure \ref{fig:nonworking}, which shows a colormap for a simulation computed with larger timestep $\mathrm{d}t = 33 \mathrm{d}t_{\text{TC}}$. One can observe that an evident ringing is produced in this case, since the time step is beyond the stable limit.

According to the analytical solution, at advanced simulation times, the interior of the ring should have a uniform temperature of $10.1667$ and outside the ring the temperature should be equal to 10. In table \ref{tabla_ring2D} we provide the accuracy measures computed for the final snapshot of our simulations ($t = 400$). We calculate the maximum and the minimum temperatures in the domain, and the error $L^1$ for the maximum temperature by comparing it with the reference value. The magnitudes of the error are within the same order, which indicates that the results obtained with the different schemes and time steps are similar with the results produced by other codes, see \cite{SHARMA2007123, 2018ApJS..234...30X} and \cite{10.1111/j.1365-2966.2012.20744.x}. 

\begin{figure}
 \centering
 \includegraphics[width = 0.5\textwidth]{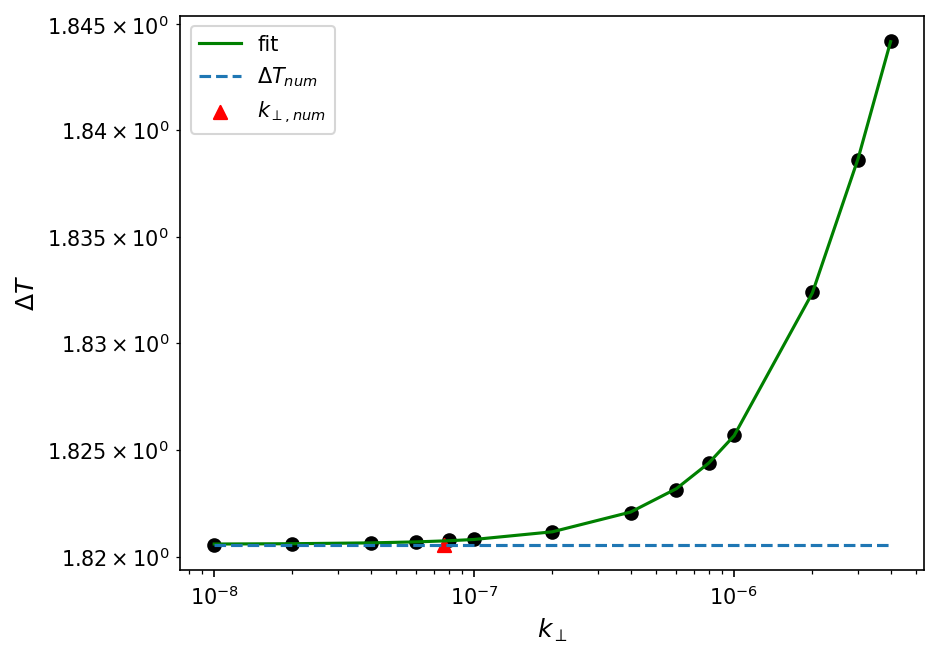}
 \caption{Graphical representation of the method used to calculate the numerical perpendicular conductivity. The dots represent the final maximum temperature minus initial maximum temperature ($\Delta T$) in the simulations with different values of $k_{\perp}$, shown at the horizontal axis. The dashed line is the difference of temperature ($\Delta T_{\text{num}}$) in the simulation with $\kappa_{\perp} = 0$. We fit the data to a hyperbolic tangent function (solid line). The  numerical perpendicular conduction coefficient  $\kappa_{\perp, \mathrm{num}}$, is estimated by locating the position where the first derivative of the fit is 100 times smaller than its maximum value. } 
 \label{fig:test_2D_numerical_perpendicular_conduction}
\end{figure}

The perpendicular numerical diffusion may cause non-negligible cross-field conduction. To estimate its value, we perform simulations with different explicit perpendicular conductivity $\kappa_{\perp}$ and calculate the difference in the maximum temperature between the final and initial stages, $\Delta T$. Similar to the calculations in \cite{2010CoPhC.181..837M}.  Figure \ref{fig:test_2D_numerical_perpendicular_conduction} shows $\Delta T$ for different values of $\kappa_\perp$. The numerical data are fit with a hyperbolic tangent function (solid line). We assign the perpendicular numerical conductivity to the value of $k_{\perp}$ at location where  the first derivative of the hyperbolic tangent function is 100 times below the maximum value of this derivative. The dashed line in Figure \ref{fig:test_2D_numerical_perpendicular_conduction} corresponds to the value of $\Delta T$ in a simulation where explicit $\kappa_{\perp}$ was set to zero ($\Delta {T}_\mathrm{num}$ = $\Delta T (\kappa_{\perp} = 0)$). It is shown in the plot to compare visually the precision of the calculation. The ratio between the numerical perpendicular conductivity and the physical parallel conductivity ($\kappa_{\perp, \mathrm{num}}/\kappa_{\|}$) for the different simulations is listed in the last column of Table \ref{tabla_ring2D}. It is of the order $\sim 10^{-5}$, which implies in one hand that we could not study the effects of anisotropic conduction in laboratory plasmas since in that case the ratio should be $\sim 10^{-9}$, otherwise perpendicular numerical diffusion will swamp the true perpendicular diffusion. On the other hand, since the ratio  it is smaller than $\sim 10^{-3}$ it should be adequate to study qualitatively the effects of anisotropic conduction on dilute astrophysical plasmas \citep{SHARMA2007123}. 

\begin{table*} 
\centering
\begin{tabular}{lccccccc}
 \hline 
 Method & N $\times$ N & $\mathrm{d}t$ & $T_{\text{max}}$  & $T_{\text{min}}$ &  $L^1$  & ${\kappa}_{\perp, \text{num}}/ \kappa_{||}$  \\ 
\hline
scheme \#1 & 200 $\times$ 200 &$\mathrm{d}t_{\text{TC}}$ & 10.166776 & 9.999981 & 7.600$\times 10^{-5}$ & 1.5755$\times 10^{-5}$ \\   
scheme \#1 & 300 $\times$ 300 &$\mathrm{d}t_{\text{TC}}$ & 10.16678  & 10.0 & 8.016$\times 10^{-5}$ & 1.2305$\times 10^{-5}$ \\   
scheme \#1 & 400 $\times$ 400 &$\mathrm{d}t_{\text{TC}}$ & 10.166782 & 10.0 & 8.188$\times 10^{-5}$ & 1.0867$\times 10^{-5}$     \\   
scheme \#1 & 500 $\times$ 500 &$\mathrm{d}t_{\text{TC}}$ & 10.166783 & 10.0          & 8.288$\times 10^{-5}$ & 1.0085$\times 10^{-5}$ \\   
\hline 
scheme \#2 & 200 $\times$ 200 &10$\mathrm{d}t_{\text{TC}}$ & 10.166772 & 9.999979 & 7.226$\times 10^{-5}$ & 1.5751$\times 10^{-5}$  \\   
scheme \#2 & 300 $\times$ 300 &10$\mathrm{d}t_{\text{TC}}$ & 10.16678  & 10.0      & 8.007$\times 10^{-5}$ & 1.2305$\times 10^{-5}$ \\   
scheme \#2 & 400 $\times$ 400 &10$\mathrm{d}t_{\text{TC}}$ & 10.16678  & 10.0      & 8.050$\times 10^{-5}$ & 1.0866$\times 10^{-5}$ \\   
scheme \#2 & 500 $\times$ 500 &10$\mathrm{d}t_{\text{TC}}$ & 10.166783 & 10.0      & 8.293$\times 10^{-5}$ & 1.0085$\times 10^{-5}$ \\ 
\hline  
scheme \#2 & 200 $\times$ 200 &20$\mathrm{d}t_{\text{TC}}$ & 10.166769 & 9.999979 & 6.941$\times 10^{-5}$ & 1.5751$\times 10^{-5}$ \\   
scheme \#2 & 300 $\times$ 300 &20$\mathrm{d}t_{\text{TC}}$ & 10.166779 & 10.0 & 7.881$\times 10^{-5}$ & 1.2305$\times 10^{-5}$     \\   
scheme \#2 & 400 $\times$ 400 &20$\mathrm{d}t_{\text{TC}}$ & 10.166781 & 10.0 & 8.120$\times 10^{-5}$ & 1.0866$\times 10^{-5}$     \\   
scheme \#2 & 500 $\times$ 500 &20$\mathrm{d}t_{\text{TC}}$ & 10.166782 & 10.0 & 8.247$\times 10^{-5}$ & 1.0085$\times 10^{-5}$     \\   
\hline 
scheme \#2 & 200 $\times$ 200 &30$\mathrm{d}t_{\text{TC}}$ & 10.166767 & 9.999981 & 6.737$\times 10^{-5}$ & 1.5751$\times 10^{-5}$ \\   
scheme \#2 & 300 $\times$ 300 &30$\mathrm{d}t_{\text{TC}}$ & 10.166778 & 10.0 & 7.755$\times 10^{-5}$ & 1.2305$\times 10^{-5}$ \\   
scheme \#2 & 400 $\times$ 400 &30$\mathrm{d}t_{\text{TC}}$ & 10.16678  & 10.0 & 8.050$\times 10^{-5}$ & 1.0866$\times 10^{-5}$ \\   
scheme \#2 & 500 $\times$ 500 &30$\mathrm{d}t_{\text{TC}}$ & 10.166782 & 10.0 & 8.202$\times 10^{-5}$ & 1.0085$\times 10^{-5}$  \\   
\hline \\
\end{tabular}\caption{Accuracy values for the two-dimensional static ring problem for different sets of simulations varying the methods used, the time steps, and the resolution. In the fourth and fifth columns, the maximum and minimum temperatures at the final stage ($t = 400$) are given. The sixth column gives the error $L^1$ for the maximum final temperature, obtained by comparing with the reference value $T = 10.1667$ . Last column provides the ratio between the numerical perpendicular conductivity and the explicit parallel conductivity of $k_{||} = 0.01$.}  \label{tabla_ring2D}  
\end{table*}

\subsection{Two-dimensional hot plate test} \label{sec:test2D}

\begin{figure*}
 \centering
 \includegraphics[height = 0.24\textheight]{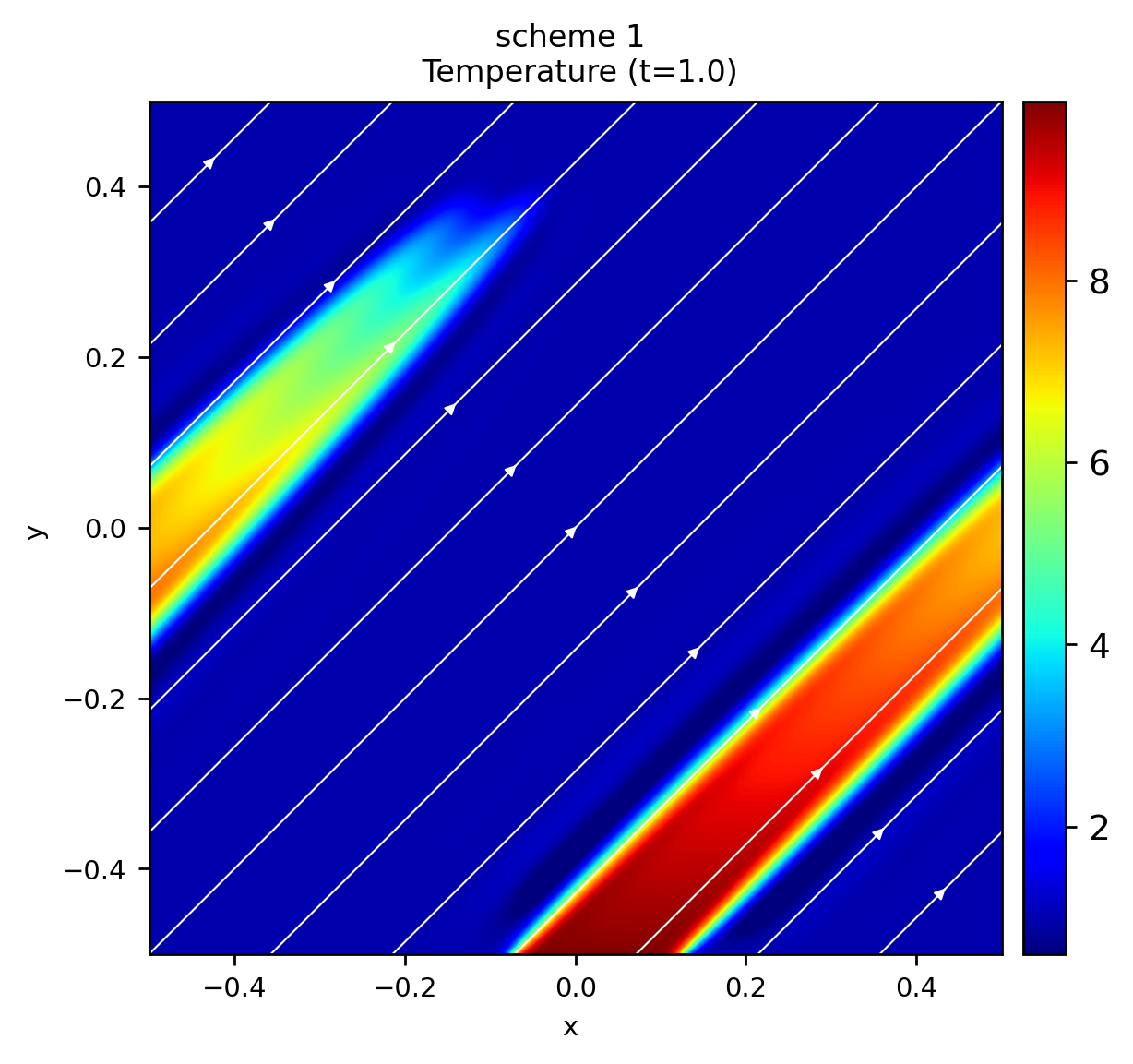}
 \includegraphics[height = 0.24\textheight]{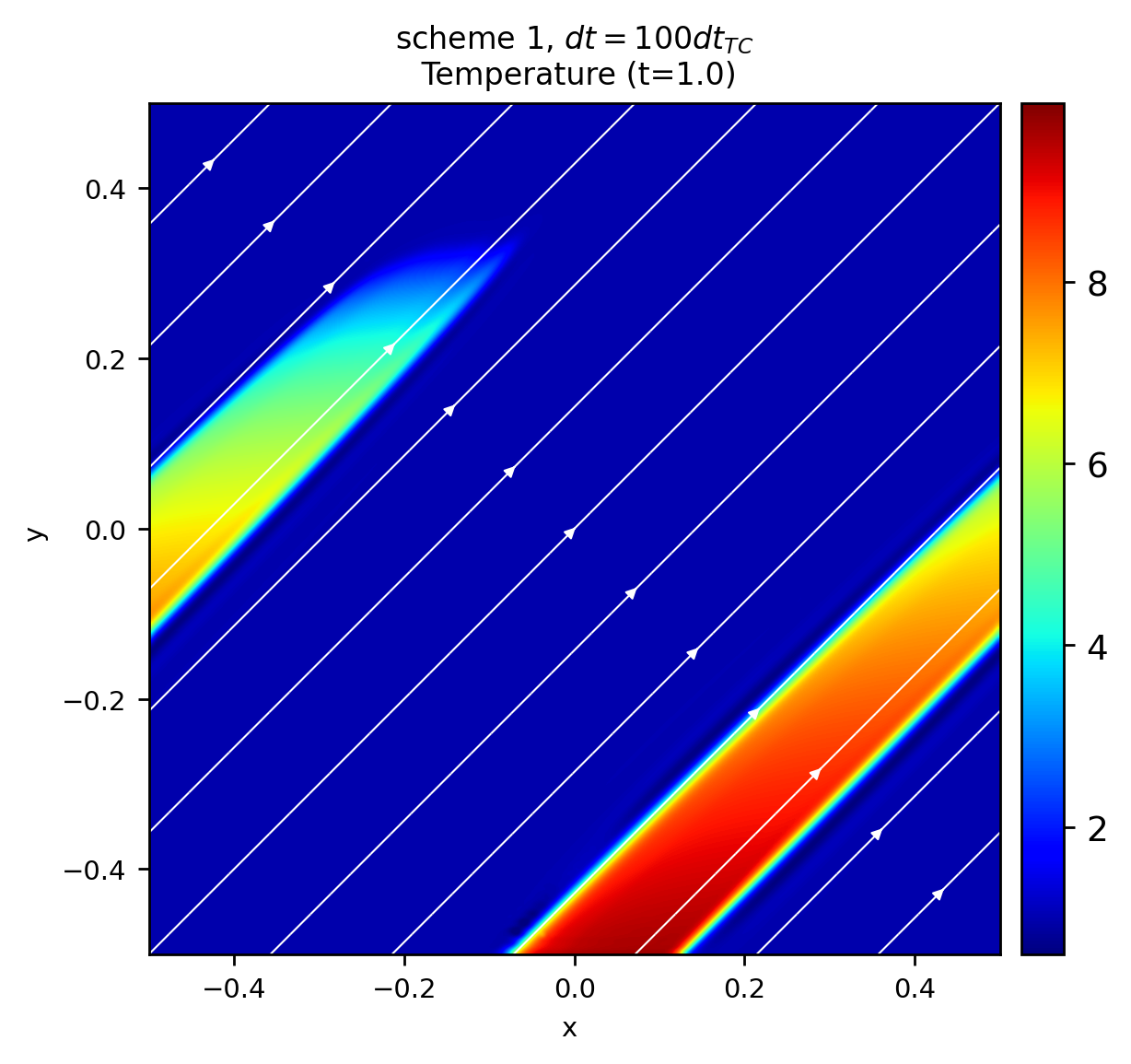}
 \includegraphics[height = 0.24\textheight]{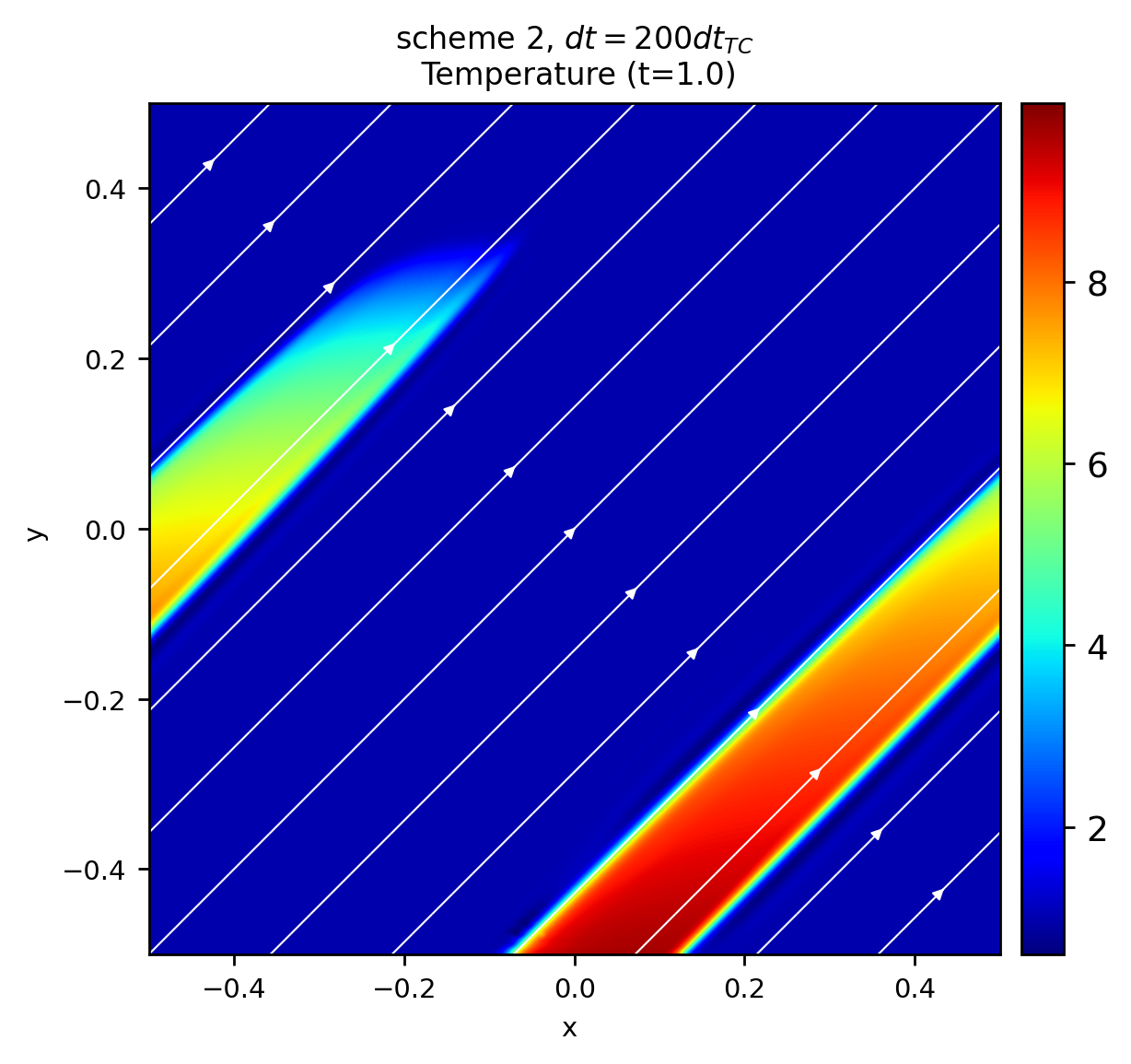}
 \includegraphics[height = 0.24\textheight]{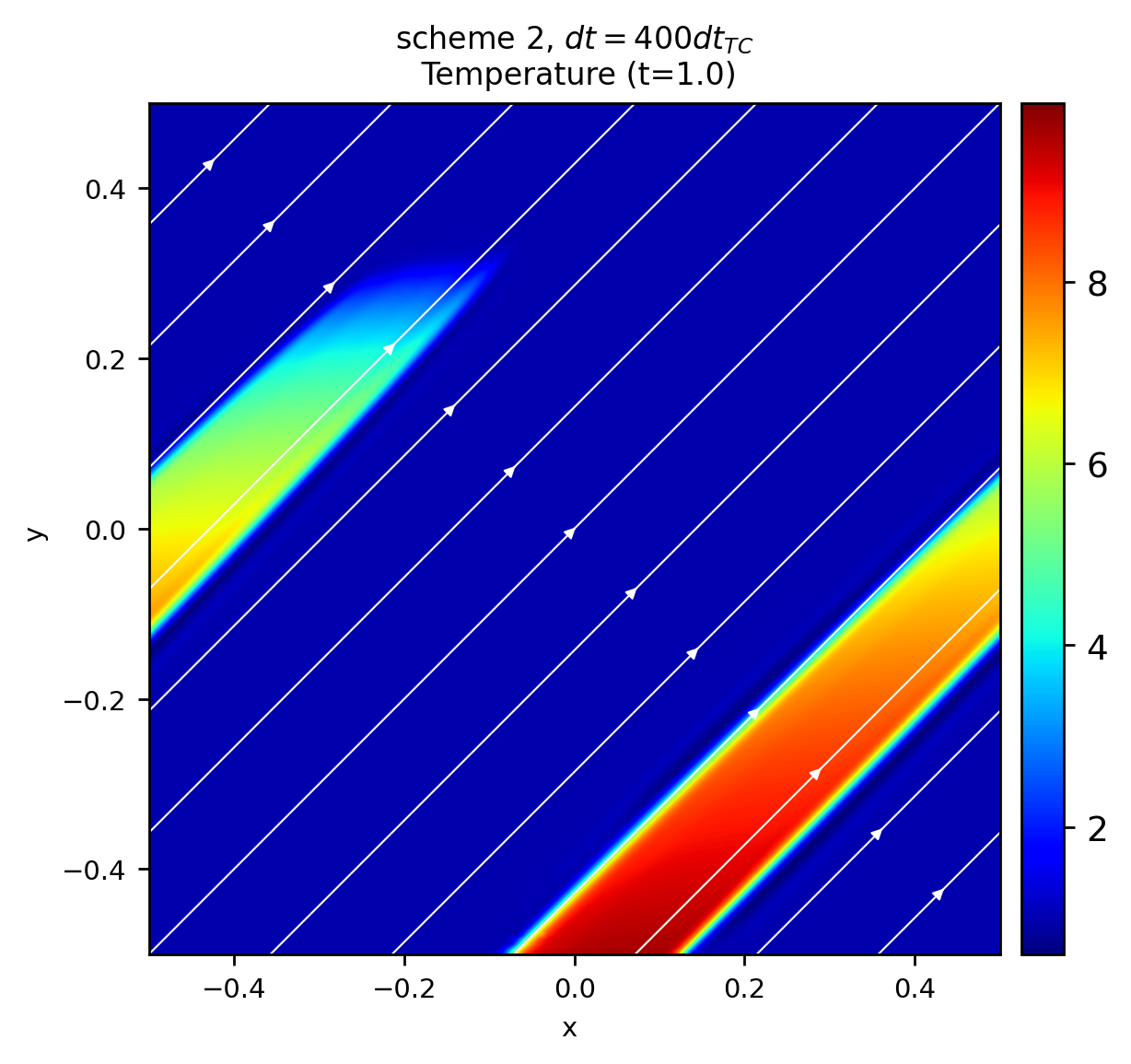}
 \includegraphics[height = 0.2\textheight]{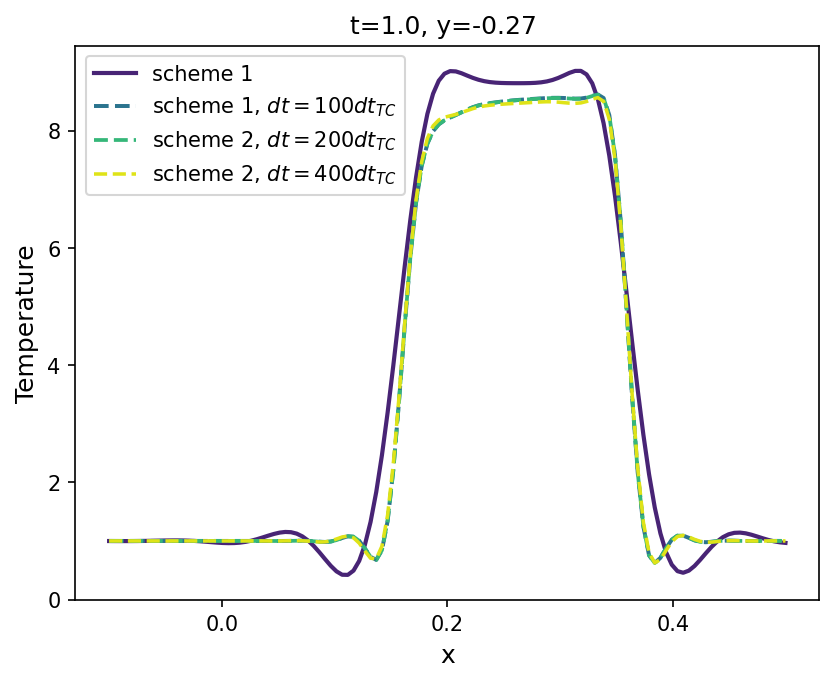}
 \includegraphics[height = 0.2\textheight]{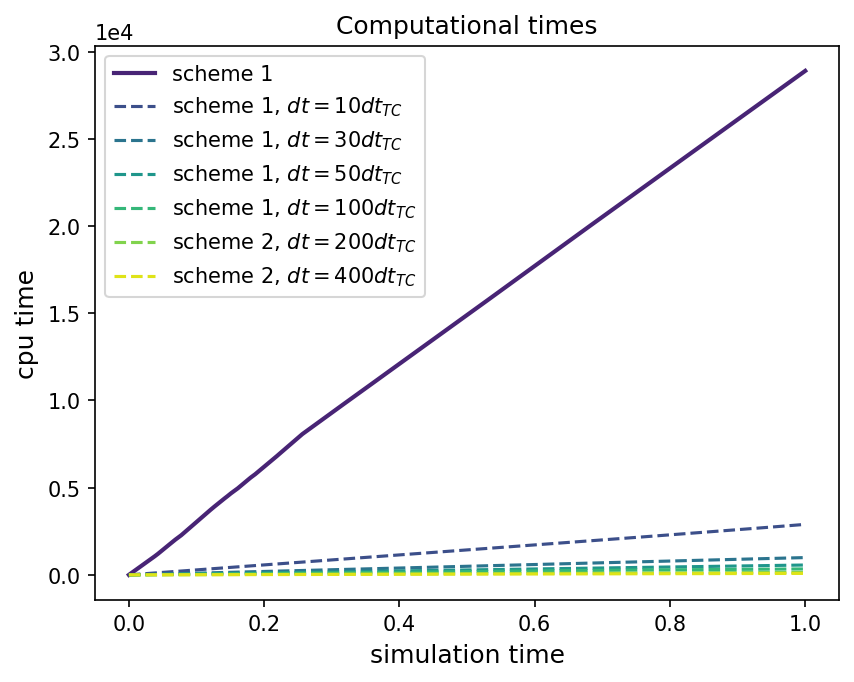}
 \caption{Left top to right middle plot: colormaps of the temperature, together with magnetic field lines, at $t=1.0$,  obtained in the different simulations, using scheme \#1 and scheme \#2 with the time steps $\mathrm{d}t = 10\mathrm{d}t_{\text{tc}}$, $\mathrm{d}t = 100\mathrm{d}t_{\text{tc}}$, $\mathrm{d}t = 200\mathrm{d}t_{\text{tc}}$ and $\mathrm{d}t = 400\mathrm{d}t_{\text{tc}}$. The bottom left plot shows a 1D cut of the temperature at $z = -0.27$, and the right bottom panel shows the simulation time for each run.} 
 \label{fig:test_2D}
\end{figure*}

A two-dimensional test designed to evaluate the anisotropic heat conduction along oblique magnetic field lines was proposed  by \cite{Jiang_2012} and reproduced by \cite{2017ApJ...844...57N}. The initial data consists of a small hot rectangular ``plate'' at the bottom of the domain.  The initial pressure is set to a homogeneous value of $p=0.1$, the system is at rest, $v_{x}=v_{y}=v_{z}=0$, and the magnetic field is set constant and inclined by a $45^{\circ}$ angle, i.e., $B_{x}=B_{y}=1, B_{z}$ $=0$. The initial density is given by
\begin{equation}
\rho=\left\{\begin{array}{l}
0.01 \text { for }|x|<0.1 \text { and } y=-0.5 \, , \\
0.1 \text { elsewhere} \, ,
\end{array}\right.
\end{equation}
in a domain extended over $[-0.5, 0.5]\times [-0.5, 0.5]$ with 200$\times$200 points. The boundary conditions are periodic in $x$, fixed to the initial values at the bottom surface at $z=-0.5$, and outflow at the top. As before, we perform simulations with both schemes. For the hyperbolic scheme, we use four times steps of $\mathrm{d}t = 10\mathrm{d} t_{\text{tc}}$, $\mathrm{d}t = 100\mathrm{d} t_{\text{tc}}$, $\mathrm{d}t = 200\mathrm{d} t_{\text{tc}}$ and $\mathrm{d}t = 500\mathrm{d} t_{\text{tc}}$. Figure \ref{fig:test_2D} shows the colormaps of the temperature and the magnetic field lines at $t=1$ for the different schemes. In all the cases the temperature is advected along the magnetic field lines at the same pace and with negligible transverse conduction.  In order to compare the differences in more detail, the left-bottom plot shows a 1D cut of the temperature at $y = -0.15$. We find that all the numerical solutions are quite similar. The right-bottom plot compares the cpu times to the solution. It again confirms that the the hyperbolic treatment produces very similar results with a great advantage in computational time. For instance, for the run where the time step is $\mathrm{d}t = 10\mathrm{d}t_{\text{tc}}$ times larger, the solution is 8.23 times faster. Similarly, the efficiency ratio for the factors 100, 200, and 500 are 76.57, 156.7 and 405.9, respectively.  In these simulations the advection time step is 1550 times larger then the conduction one. An example of an unstable simulation for this test is given in the bottom left panel of Figure \ref{fig:nonworking}. It displays a colormap in a simulation using a time step beyond the stability limit ($\mathrm{d}t = 500\mathrm{d}t_{\text{tc}}$), a clear ringing is present from an early stage. 

\subsection{Three-dimensional Static Ring} \label{sec:ring3d}

A 3D generalization of the static ring problem was proposed by \cite{2018ApJS..234...30X}. Starting from the set up described in section \ref{sec:ring2d}, we extend the hot patch over a width of $0.4$ in $z$ direction. The whole system, including the magnetic field, is rotated by $\pi/4$ around the $x$-axis and by $\pi/4$ around the $z$-axis. This rotation causes a misalignment between the field and the coordinate axes, so that the thermal flux evolves in the three dimensions. We run four different simulations in the numerical box of $[-1,1] \times [-1,1] \times [-1,1]$ with $200 \times 200 \times 200$ points. The first one is with the scheme \#1, and remaining three are with scheme \#2 with time steps  $\mathrm{d}t = 10 \mathrm{d} t_{\text{TC}}$,  $\mathrm{d} t = 20 \mathrm{d} t_{\text{TC}}$,  $\mathrm{d}t = 30 \mathrm{d}t_{\text{TC}}$ and  $\mathrm{d}t = 40 \mathrm{d}t_{\text{TC}}$.

\begin{figure*}
 \centering
 \includegraphics[height = 0.26\textheight]{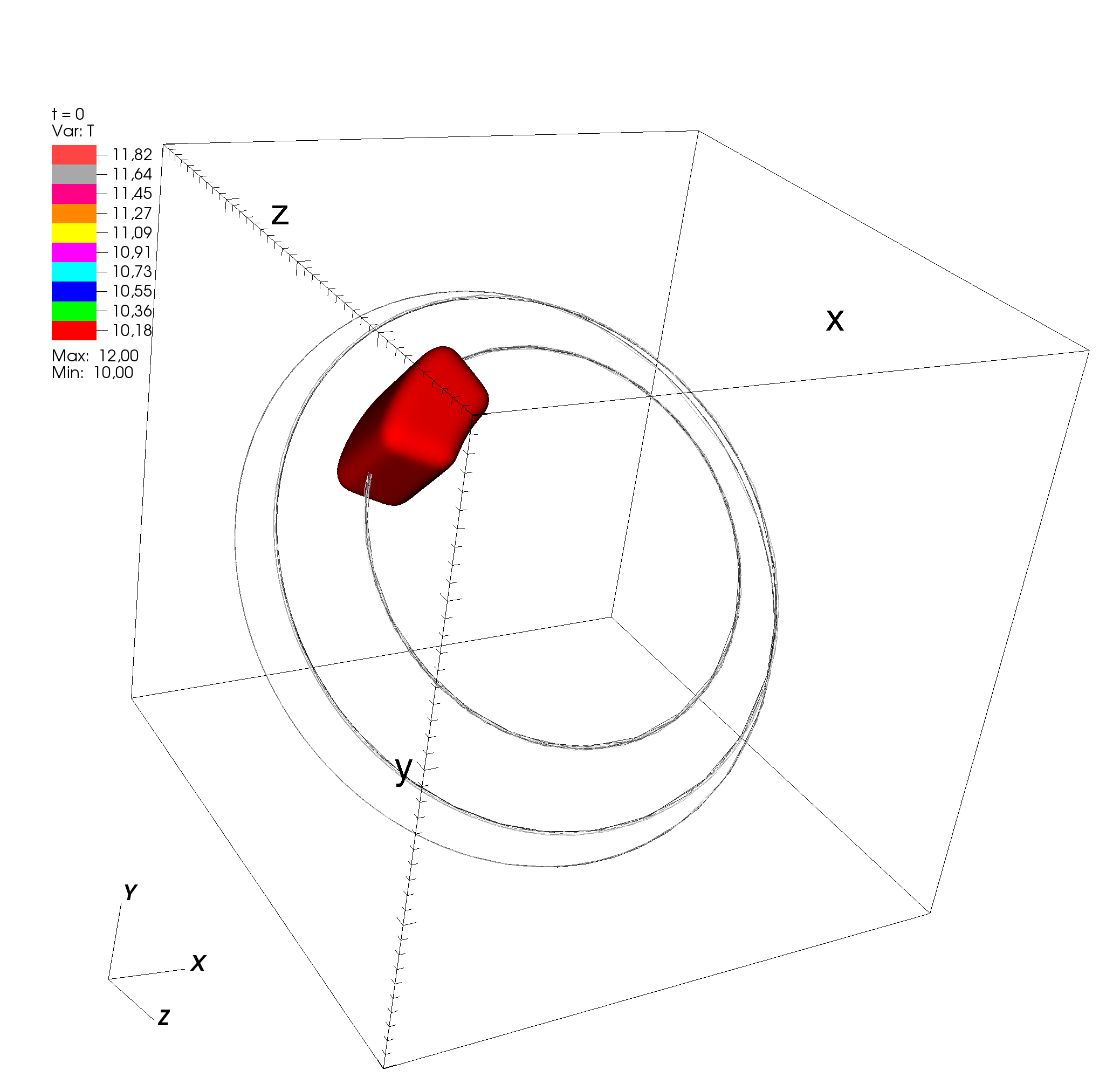}
 \includegraphics[height = 0.26\textheight]{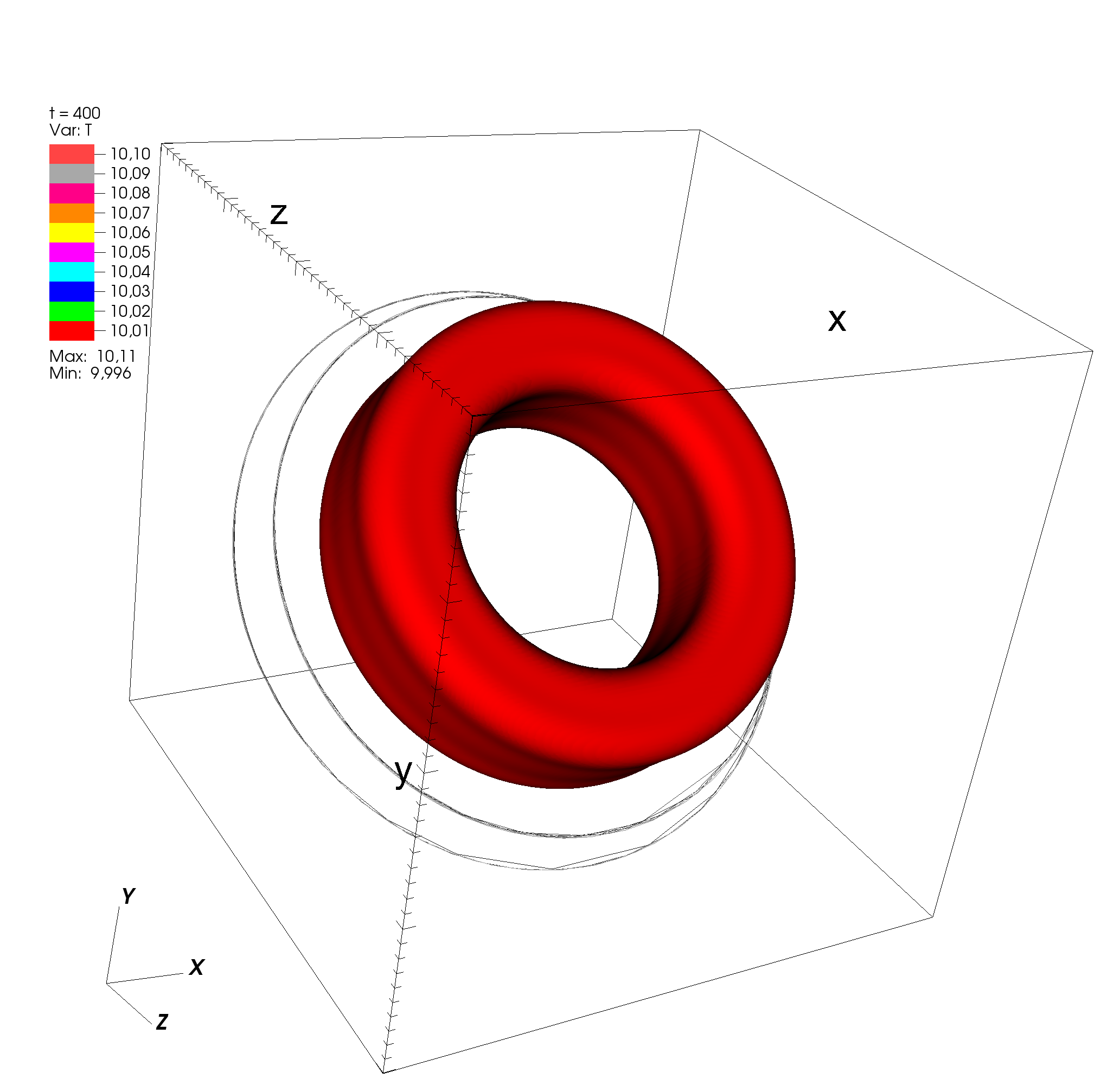}\\ 
 \includegraphics[height = 0.21\textheight]{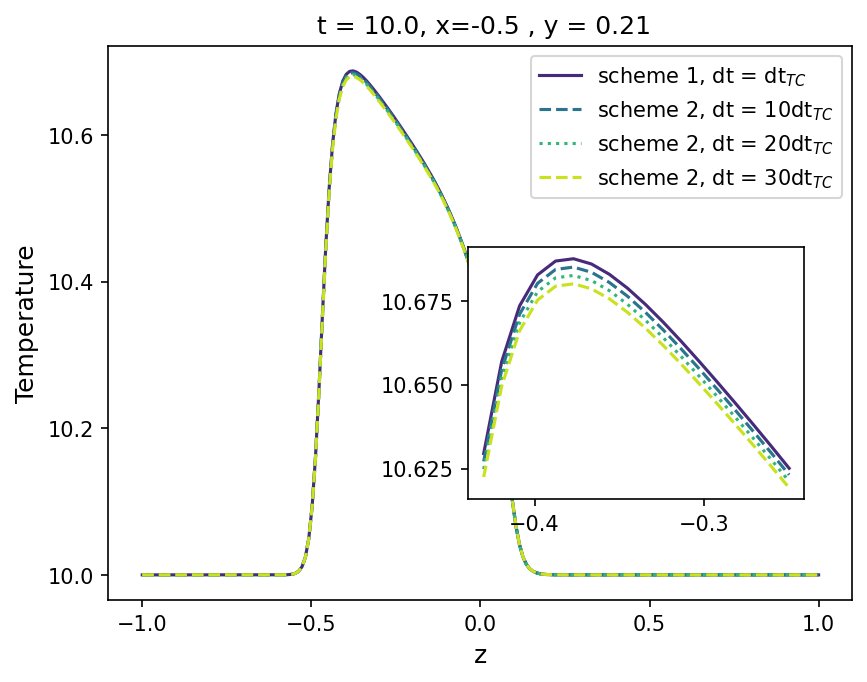}
 \includegraphics[height = 0.21\textheight]{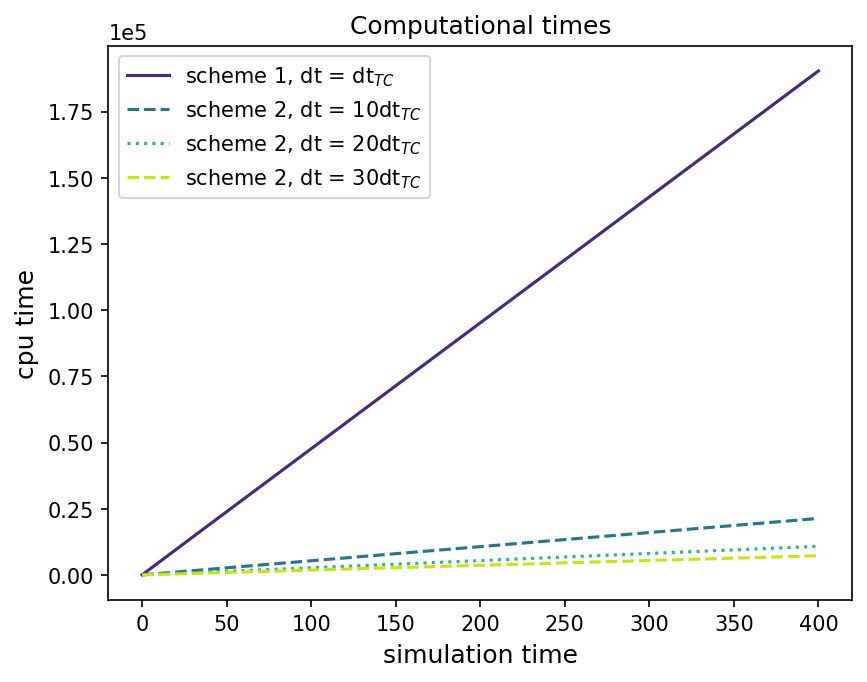}\\ 
 \caption{Results for the three-dimensional static ring problem. The top panels correspond to the contour of temperature at the initial and final stages, together with the magnetic field lines, obtained with the simulation with scheme \#1. Bottom left panel shows a comparison of one snapshot of the temperature at $x = -0.5$ and $y=0.21$ at $t=10$ for different simulations using scheme \#2 and different time steps. In the bottom right panel, a graphical representation of the computational times of the different runs is given.}
 \label{fig:test_3D}
 \end{figure*}

\begin{figure*} 
 \centering
\includegraphics[height = 0.2\textheight]{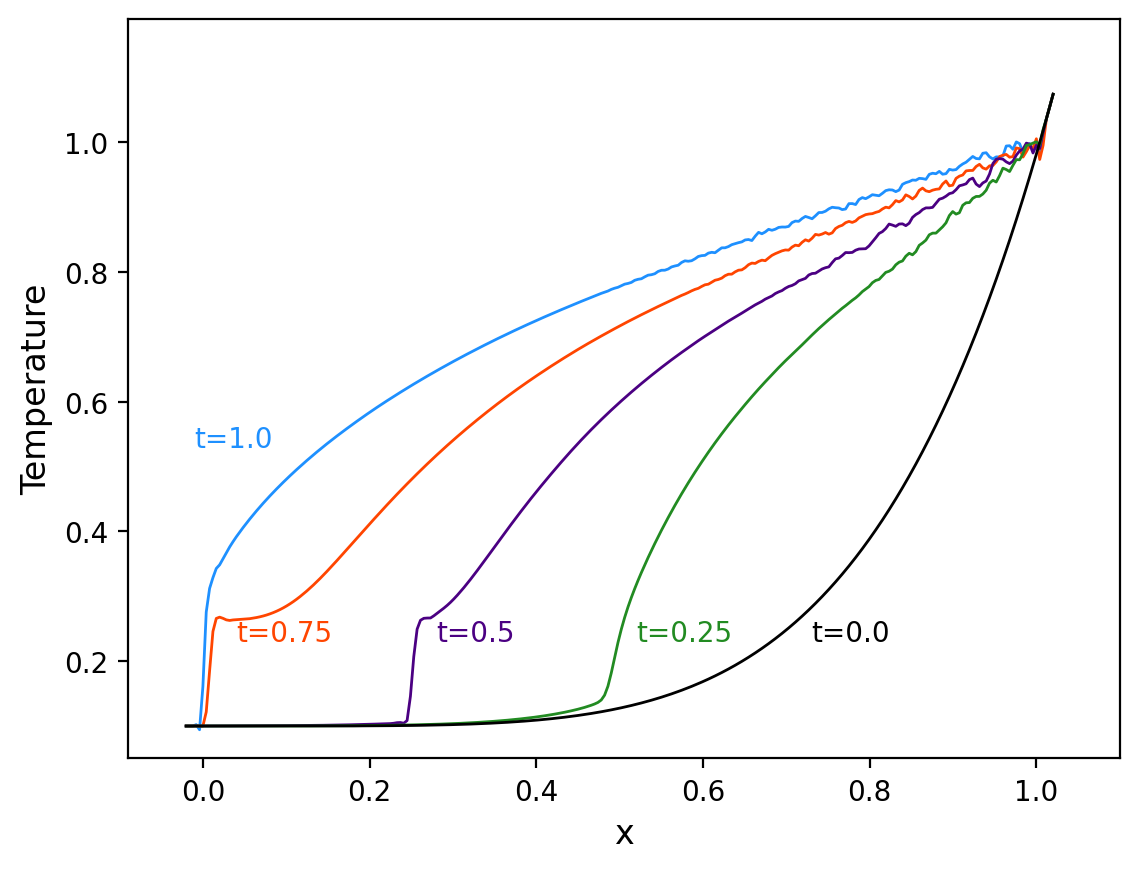}
 \includegraphics[height = 0.2\textheight]{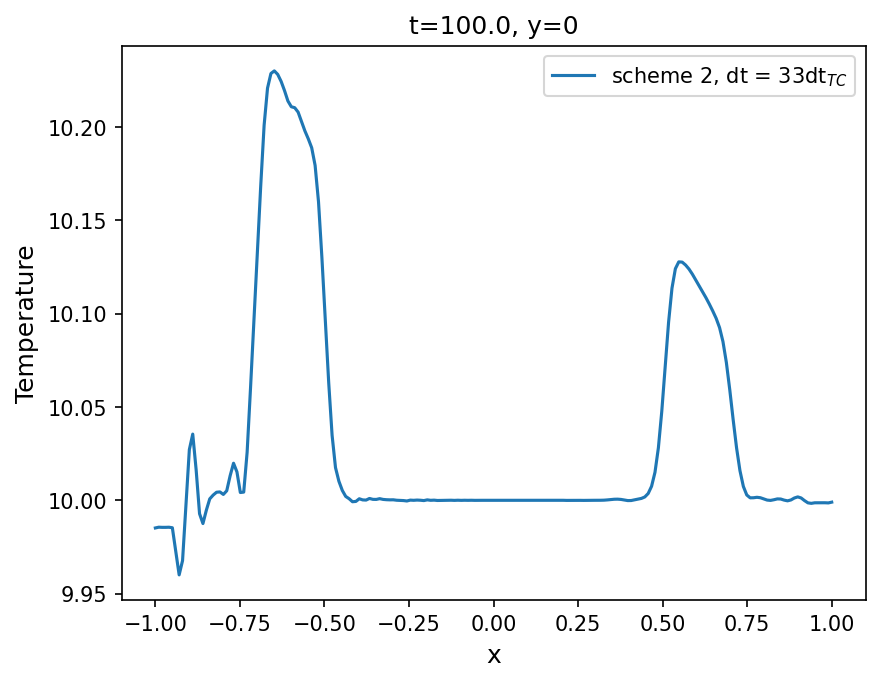}
 \includegraphics[height = 0.2\textheight]{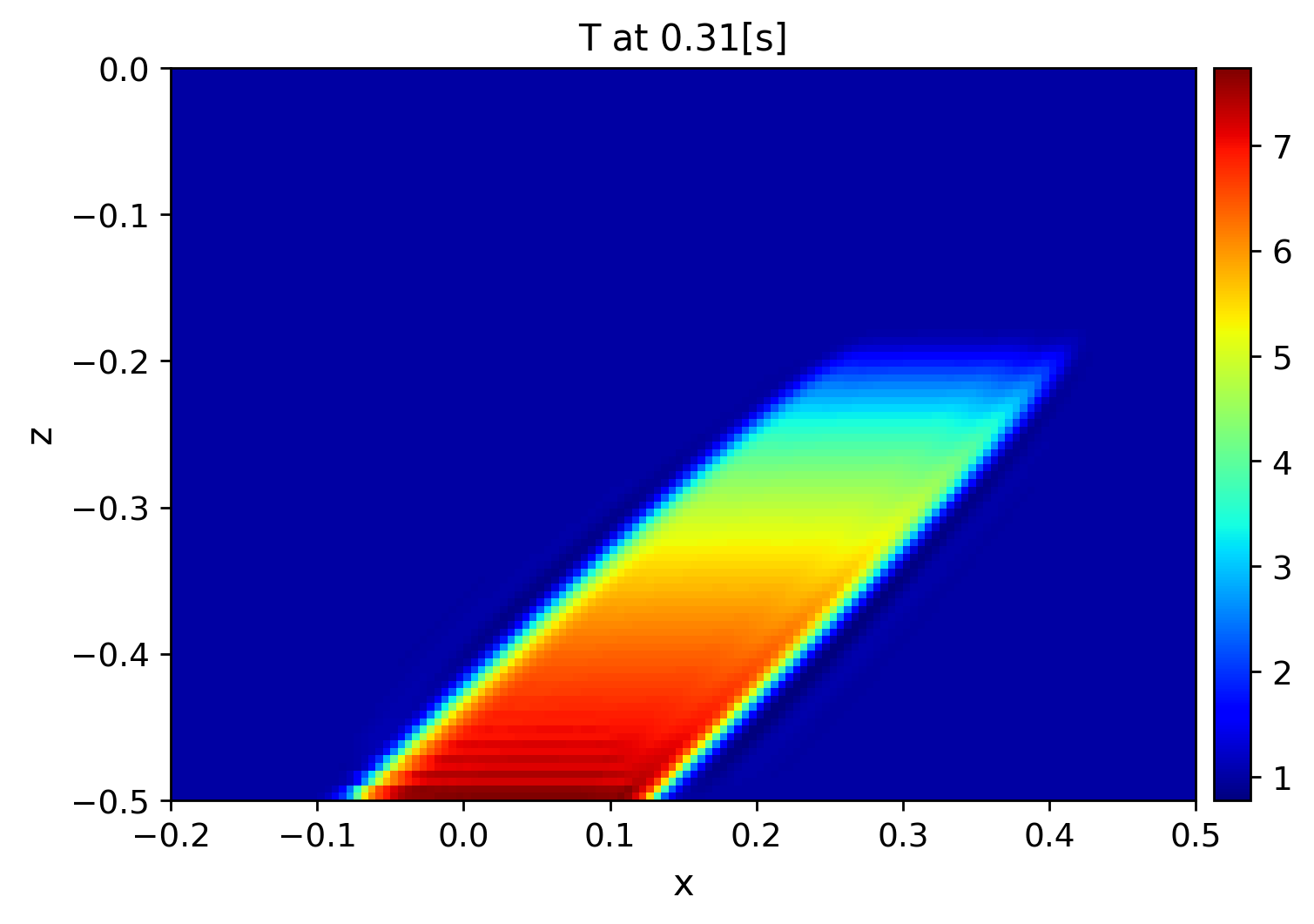}
  \includegraphics[height = 0.2\textheight]{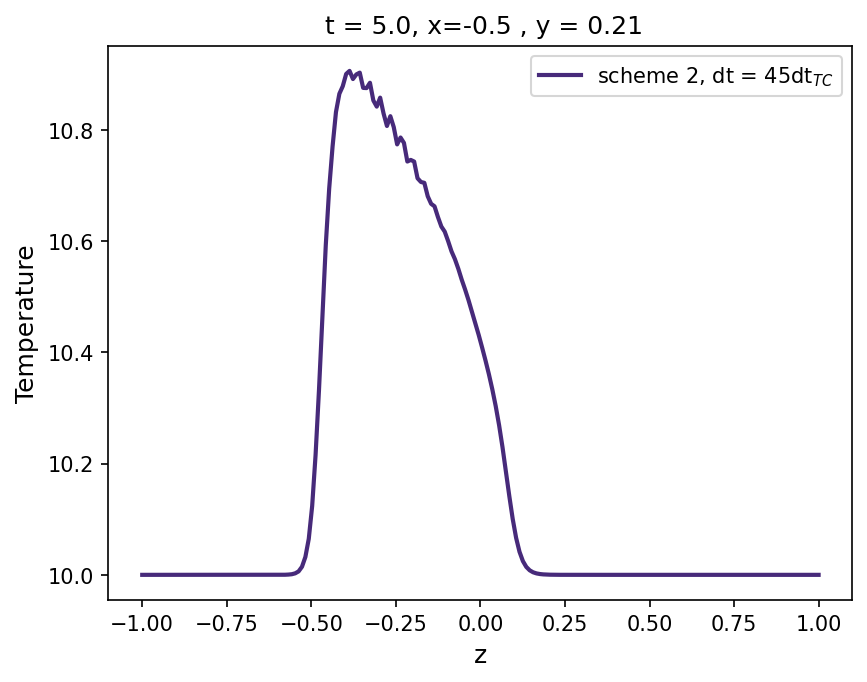}
 \caption{Unstable examples of the tests using scheme \#2 with time steps beyond the stability limit. Top left shows the temperature evolution in the one-dimensional test with $\mathrm{d}t = 200 \mathrm{d}t_{\text{TC}}$. Top right presents a temperature colormap for the static ring problem using $\mathrm{d}t = 33 \mathrm{d}t_{\text{TC}}$ and $t = 100$. The bottom left panel shows a temperature colormap of the temperature for the hotplate test, using a time step $\mathrm{d}t = 500\mathrm{d}t_{\text{tc}}$. Bottom right panel shows a contour of temperature in the three-dimensional static ring problem in an early stage, $t = 5$ s, using a time step $\mathrm{d}t = 45\mathrm{d}t_{\text{tc}}$.} 
 \label{fig:nonworking}
\end{figure*}

Top panels of Figure \ref{fig:test_3D} present a contour of the temperature together with magnetic field lines at the initial and last stage of the simulation with scheme \#1. Bottom left panel allows us to compare the results between the different runs showing the temperature at $x = -0.5$ and $y=0.21$ at the final stage $t=400$. One can observe that there is a slight difference in the temperature profiles. In particular, the runs with the scheme \#2 with larger time step ($\mathrm{d}t = 20\mathrm{d}t_{\text{tc}}$ and $\mathrm{d}t = 30\mathrm{d}t_{\text{tc}}$) exhibit less perpendicular dissipation since the maximum values of the temperature in those cases are bigger. Additionally, the bottom right panel of Figure \ref{fig:test_3D} displays the computational times for the simulations, showing the big advantage in computational times of the hyperbolic treatment. In these simulations the advection time step is 150 times larger then the conduction one. For the simulation where the time step is $\mathrm{d}t = 10\mathrm{d}t_{\text{tc}}$, the efficiency ratio is 4.9. Similarly, the efficiency ratio for the factors 20 and 30, are of 9.45, and 14.6 respectively. An example of a unstable simulation using scheme \#2 is presented in bottom right panel of Figure \ref{fig:nonworking}, which shows a ringing produced by the large time step, in this case, $\mathrm{d}t = 45\mathrm{d}t_{\text{tc}}$. 

Table \ref{tabla_ring3D} presents the maximum temperature reached at the final stage and provides the $L^1$ error computed by comparing this maximum temperature with the theoretical value $T =$ 10.1667. The values of the error are quite similar in the different simulations and with the results obtained in the 2D case, proving that the code resolves with sufficient accuracy three-dimensional thermal conduction problems with complex magnetic field configurations. 

\begin{table} 
\centering
\begin{tabular}{lccccc}
 \hline 
 Method & $\mathrm{d}t$ & $T_{\text{max}}$ & $T_{\text{min}}$ & $L^1$  &   \\ 
\hline
scheme \#1 &   $\mathrm{d}t_{\text{TC}}$ & 10.166782  & 9.999995 & 8.230$\times 10^{-5}$ \\    
scheme \#2 & 10$\mathrm{d}t_{\text{TC}}$ & 10.166781 & 9.999995 & 8.055$\times 10^{-5}$   \\    
scheme \#2 & 20$\mathrm{d}t_{\text{TC}}$ & 10.166779 & 9.999995 & 7.885$\times 10^{-5}$   \\    
scheme \#2 & 30$\mathrm{d}t_{\text{TC}}$ & 10.166777 & 9.999995 & 7.715$\times 10^{-5}$   \\    
 \hline      
 \end{tabular}\caption{Maximum temperature and error $L^1$ for the three-dimensional static ring problem from section \ref{sec:ring3d}.}  \label{tabla_ring3D}  
 \end{table} 


\section{Thermal conduction in the solar atmosphere} \label{sec:braginskii}

In a magnetized plasma, the efficiency of the heat conduction depends on the magnetic field direction. It is therefore convenient to decompose the conductivity vector using the projections into the parallel and perpendicular to the field directions,
\begin{eqnarray}
{\bf q}= -\kappa_{\|}\nabla_{\|}T - \kappa_{\perp}\nabla_{\perp}T +\kappa_{\times} {\bf \hat{b}} \times \nabla_{\perp} T \, , \label{eq:q_decomposition}
\end{eqnarray}
where $\nabla_{\|} = {\bf \hat{b} ({\bf \hat{b}} \cdot \nabla) }$ gives the parallel projection to the field, $\nabla_{\perp} = \nabla - \nabla_{\|}$ gives the projection in the perpendicular direction, and the last term is the projection in the transverse direction (second perpendicular direction to the magnetic field). For example, similar decomposition was recently used in \citet{Peters_manuscript}. \cite{1965RvPP....1..205B} deduced the following general expressions for the conductivity coefficients used in the expressions of the heat conduction vector for electrons  
\begin{eqnarray}
\kappa_{\|}^{e} &=&  3.1616\frac{ k_B p_{e}}{\nu_{ei} m_{e}} \, , \label{eq:brag_par_e} \\
 \kappa_{\perp}^{e} &=& \frac{k_B p_{e}}{\nu_{ei} m_{e}} \frac{4.664 x_e^{2}+11.92}{x_e^{4}+14.79 x_e^{2}+3.77} \, , \label{eq:brag_perp_e}  \\
\kappa_{\times}^{e} &=& \frac{k_B p_{e}}{\nu_{ei} m_{e}} x_e \frac{\frac{5}{2} x_e^{2}+21.67}{x_e^{4}+14.79 x_e^{2}+3.77} \, , \label{eq:brag_cross_e}
\end{eqnarray}
and for ions,
\begin{eqnarray}
\kappa_{\|}^{i} &=& 3.906 \frac{k_B p_{i}}{\nu_{ii} m_{i}}  \, ,  \label{eq:brag_par_i} \\ 
\kappa_{\perp}^{i} &=& \frac{k_B p_{i}}{\nu_{ii} m_{i}} \frac{2 x_i^{2}+2.645}{x_i^{4}+2.70 x_i^{2}+0.677}  \label{eq:brag_perp_i} \, , \\ 
\kappa_{\times}^{i} &=& \frac{k_B p_{i}}{\nu_{ii} m_{i}} x_i \frac{\frac{5}{2} x_i^{2}+4.65}{x_i^{4}+2.70 x_i^{2}+0.677} \, , \label{eq:brag_cross_i}
\end{eqnarray}
where the subindices and upper indices $e$ and $i$ refer to electrons and ions. The pressure and mass of each specie are $p_e$, $p_i$, $m_e$ and $m_i$. The collisional frequency of ion-ion collisions is $\nu_{ii}$ and of electron-ion collissions is $\nu_{ei}$. The quantities $x_i$ and $x_e$ represent the ratio between cyclotron frequency ($\Omega$) and collision frequency ($\nu$) for electrons and ions, respectively
\begin{eqnarray}
x_e = \frac{\Omega_e}{\nu_{ei}}; \,\,\, x_i = \frac{\Omega_i}{\nu_{ii}} \, , 
\end{eqnarray}
where
\begin{eqnarray} \label{eq:omegae}
\Omega_e = - \frac{|e|B}{m_e}; \,\,\, \nu_{ei} = 3.7\times 10^{-6} \frac{ \ln(\Lambda) n_e}{T_e^{3/2} }; \, \, \, p_e =  n_e k_B T_e \, , \\
\label{eq:omegai}
\Omega_i = \frac{|e|Z_iB}{m_i}; \,\,\, \nu_{ii} = 6\times 10^{-8} \frac{ \ln(\Lambda) n_i Z_i^4}{T_i^{3/2}}; \, \, \, p_i = n_i k_B  T_i \, , 
\end{eqnarray}
$e$ is the electron charge, $Z_i$ is the charge of the ion, $n_e$ and $n_i$ the electron and ion number density. All the definitions in SI units.
The Coulomb logarithm could be approximated by
\begin{equation}
\ln(\Lambda) = 23.4 - 1.15 \log_{10}(n_e) + 3.45\log_{10}T_e \, ,    
\end{equation}
for $T_e<50$ eV, with $n_e$ given in cm$^{-3}$ and $T_e$ in electronvolts. It follows from these expressions that the case of null magnetic field the heat flux becomes isotropic ($\vec{q} = -k_{||} \nabla T$) since $x_e = x_i = 0$ and $k_\perp^i = \kappa_{\|}^e$.  

The parallel conductivity coefficients given in (\ref{eq:brag_par_e}) and (\ref{eq:brag_par_i}) have a dependence on temperature as $\sim T^{5/2}$ (this follows from the dependence on temperature of the pressure and the collision frequencies, see Eqs. \ref{eq:omegae}, \ref{eq:omegai}). The expression for the electron heat flux is the same as given by Spitzer, after exchanging the constant 3.203 for 3.1616 in Eq. \ref{eq:brag_par_e}). This dependence on temperature is specially problematic due to the high temperatures of the corona, when solving the thermal conductivity with an explicit integration method.  We propose to combine the schemes \#1 and \#2 by solving the energy equation, Eq. \ref{eq:E_q}, where the heat flux is given by 
\begin{eqnarray}
{\bf q}= - q_{\|} {\bf \hat{b}}  - \kappa_{\perp}\nabla_{\perp}T +\kappa_{\times} {\bf \hat{b}} \times \nabla_{\perp} T \, ,
\end{eqnarray}
with the hyperbolic equation for the evolution of the parallel heat flux component $q_{\|}$,
\begin{eqnarray}
\frac{\partial q_{\|}}{\partial t} = \frac{1}{\tau} \left(  f_{\text{sat}} \kappa_{\|} \left( {\bf \hat{b}} \cdot \nabla \right) T + q_{\|} \right) \, .
\end{eqnarray}
In the latter expression, $\kappa_{\|}$ is the conductivity coefficient in parallel direction to the magnetic field 
\begin{eqnarray}
\kappa_{\|} = \kappa_{\|}^e + \kappa_{\|}^i \, ,
\end{eqnarray}
 together which includes the contributions from electrons $\kappa_{\|}^e$ (\ref{eq:brag_par_e}) and ions $\kappa_{\|}^i$ (\ref{eq:brag_par_i}), assuming $T_e = T_i = T$, $n_e = n_i$ and $Z_i = 1$. Accordingly, $\kappa_{\perp}$ is the conductivity coefficient in the perpendicular direction to the magnetic field
\begin{eqnarray}
\kappa_{\perp} = \kappa_{\perp}^e + \kappa_{\perp}^i \, ,
\end{eqnarray}
which includes the contributions from electrons $\kappa_{\perp}^e$ (\ref{eq:brag_perp_e}) and ions $\kappa_{\perp}^i$ (\ref{eq:brag_perp_i});
$\kappa_{\times}$ is the transversal conductivity coefficient, perpendicular to both the magnetic field and the temperature gradient
\begin{eqnarray}
\kappa_{\times} = \kappa_{\times}^e + \kappa_{\times}^i \, ,
\end{eqnarray} 
and includes the contributions from electrons $\kappa_{\times}^e$ (\ref{eq:brag_cross_e}) and ions $\kappa_{\times}^i$ (\ref{eq:brag_cross_i}). 

Since $q_{\|}$ is an independent variable, its initial value should be calculated like 
\begin{eqnarray}
q_{\|}(t=0) =  f_{\text{sat}} \kappa_{\|} \left( {\bf \hat{b}} \cdot \nabla \right) T \, . 
\end{eqnarray}

\begin{figure*}
    \centering
    \includegraphics[height = 0.21\textheight]{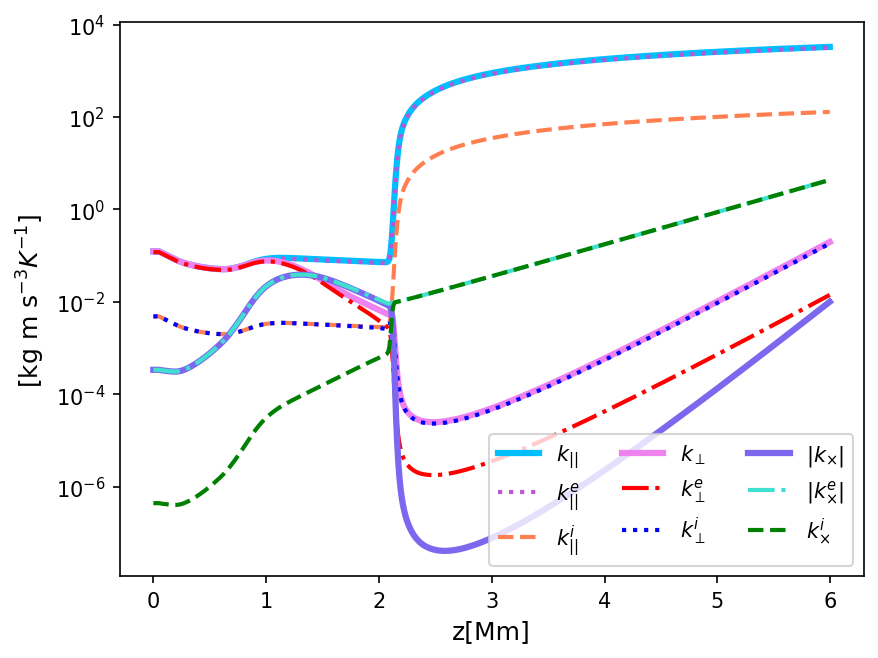} 
    \includegraphics[height = 0.21\textheight]{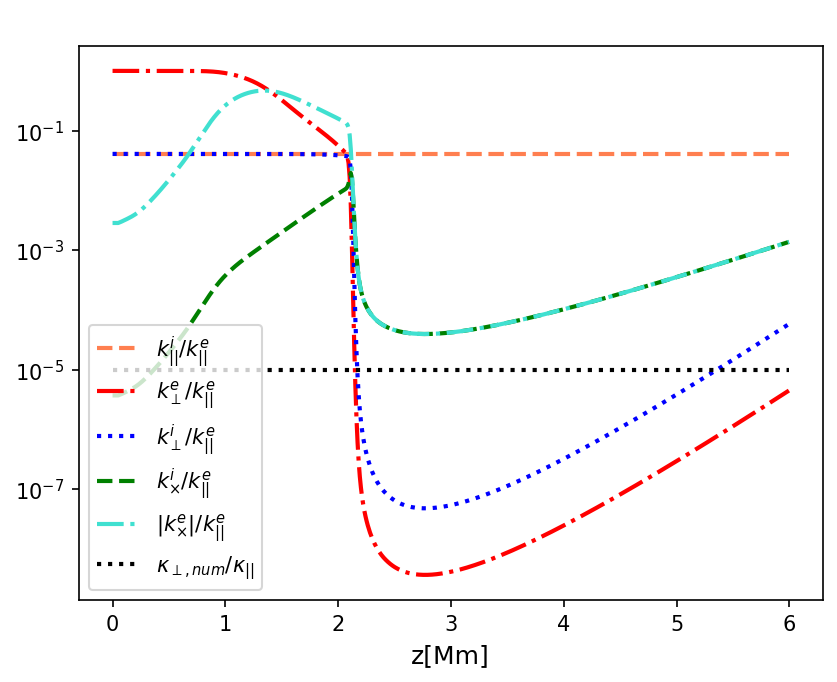} \\ 
    \caption{Thermal conductivity coefficients for a stratified solar atmosphere with the temperature model of  \cite{Vernazza_temperature_profile} extended to the corona and a  magnetic field for the quiet sun given by Eq. \ref{eq_B_quiet}. Left panel: different components for ions and electrons. Right panel: ratios between different conductivity coefficients and the parallel electron conductivity coefficient.}
    \label{fig:stratification}
 \end{figure*}

The model above takes into account six different conductivity coefficients. It is of interest to compare the magnitude of the terms as a function of height and magnetisation of the solar stratified atmosphere. For that, the left panel of Figure \ref{fig:stratification} presents those conductivities as functions of height, considering a stratified atmosphere with the temperature model of \cite{Vernazza_temperature_profile} extended to the corona.  We adopted the following approximate dependence for the magnetic field, appropriate for a quiet solar region, 
\begin{equation}
B_z = (100 \textrm{G}) \exp{(-z/(6 \textrm{Mm}))} \, .  \label{eq_B_quiet}
\end{equation}
The results will of course vary if a different strength and stratification for the magnetic field is adopted. However, the conclusions below will remain qualitatively similar, since the dependence of the conductivity coefficients on the thermodynamic parameters  is much stronger.

According to the left panel of Figure \ref{fig:stratification}, the parallel conductivity $k_{\|} = k_{\|}^e + k_{\|}^i$ is the most important one at all heights. Although in the photosphere it is equal in magnitude to the perpendicular one $k_{\perp} = k_{\perp}^e + k_{\perp}^i$, which has its largest values in the lower atmosphere. The biggest contribution of the transverse conductivity $k_{\times} = k_{\times}^e + k_{\times}^i$ is before the transition region, after, it drops drastically and increases smoothly according to height. Now, to determine specifically the contributions of each conductivity due to electrons or ions, the right panel of Figure \ref{fig:stratification} provides the ratios $\kappa_{\|}^i/\kappa_{\|}^e$, $\kappa_{\perp}^e/\kappa_{\|}^e$, $\kappa_{\perp}^i/\kappa_{\|}^e$, $\kappa_{\times}^i/\kappa_{\|}^e$ and $\kappa_{\times}^e/\kappa_{\|}^e$. The ratios are computed over the parallel conductivity of electrons ($k_{\|}^e$) which is the most dominant at almost all heights. In this plot, we can see that below the transition region ($z = 2.1$ Mm), the perpendicular conductivity of electrons ($\kappa_{\perp}^e$) is the second more important and is larger than all the conductivities from ions. The parallel conductivity of ions ($\kappa_{\|}^i$) is proportional to the parallel conductivity of electrons ($\kappa_{\|}^e$) and is around one order of magnitude smaller. This might be evident since we are assuming equal temperatures for both electrons and ions to plot this figures. However, it shows that in the corona, the parallel conductivity of ions ($\kappa_{\|}^i$) is the second most dominant.  Additionally, both transverse conductivities for ions  and electrons ($\kappa_{\times}^i$ and $\kappa_{\times}^e$) may contribute below the transition region.  At other heights, the other components do not contribute much in this atmosphere model since they are smaller than the numerical perpendicular conductivity ratio of the code ($\sim 10^-5$), which was found in Section \ref{sec:ring2d}.

\subsection{2D test with a realistic solar temperature model} \label{sec:hottop} 

We present an adaptation of the hot plate test, discussed above in Section \ref{sec:test2D}, to a realistic density and temperature profile for the solar atmosphere. This is done to verify the performance of our implementation of the thermal conductivity using realistic high temperatures of the solar corona. Including conductivity in such model in the parabolic treatment decreases the time step three orders of magnitude in comparison to the advective time steps. 

The initial state consists of a stratified temperature profile of \cite{Vernazza_temperature_profile} smoothly joint to a constant-temperature corona at $1.2\times 10^6$ K. The pressure and densities are obtained from the hydrostatic equilibrium equation and the ideal gas law
\begin{eqnarray}
p(z) &=& p_0 \exp \left(-\frac{m_p g}{k_{\mathrm{B}}} \int_{z_0}^{z} \frac{\mathrm{d} \tilde{z}}{T(\tilde{z})}\right)  \, , \\
\rho(z) &=& \frac{m_p }{k_{\mathrm{B}}} \frac{p(z)}{T(z)} \, ,
\end{eqnarray}
where $m_{\mathrm{p}}$ is the proton mass, $k_{\mathrm{B}}$ is the Boltzmann constant, $p_{0} = 0.1$ Pa and $z_0 = $ 10 Mm is a reference height. The magnetic field strength is taken from Eq. \ref{eq_B_quiet}, and we set it to be inclined by $45^{\circ}$ degrees. The numerical domain extends over $[-1,1]\times[0,7]$ Mm with 400$\times$1400 points. We set a hot region at the top boundary between $x = -0.25$ Mm and $x = $ 0.25 Mm with a temperature $10\%$ higher than the equilibrium. The boundary conditions are periodic in the $x$ direction and fixed to the initial values at the top and bottom.  

\begin{figure*}
\centering
\includegraphics[height = 0.4\textheight]{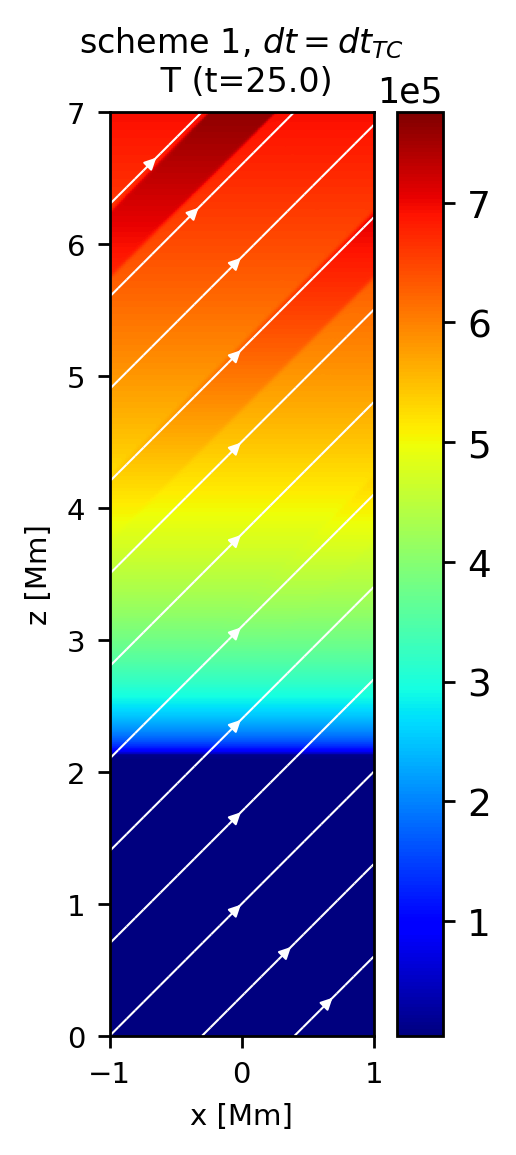}
\includegraphics[height = 0.4\textheight]{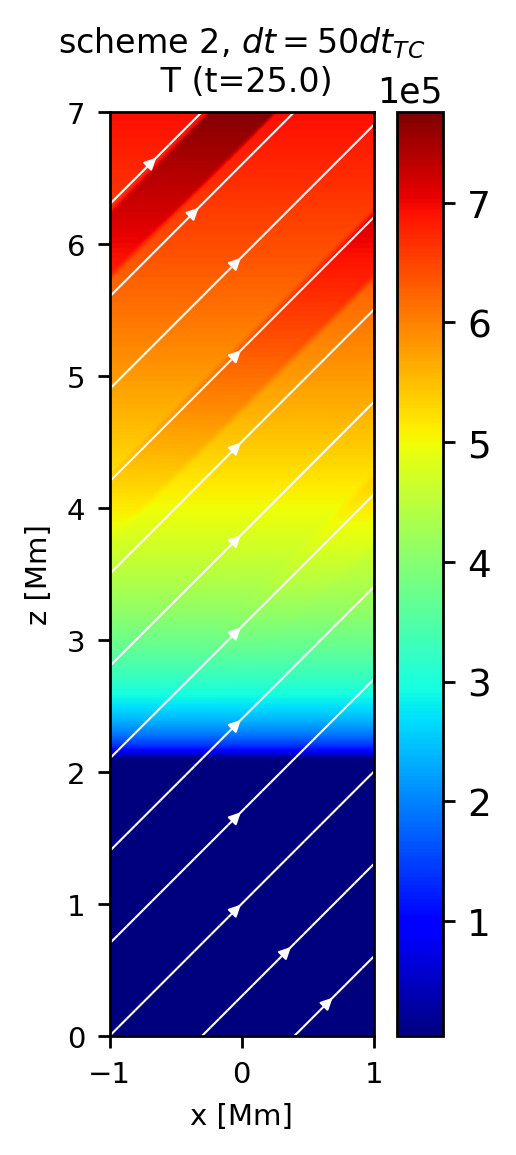}
\includegraphics[height = 0.4\textheight]{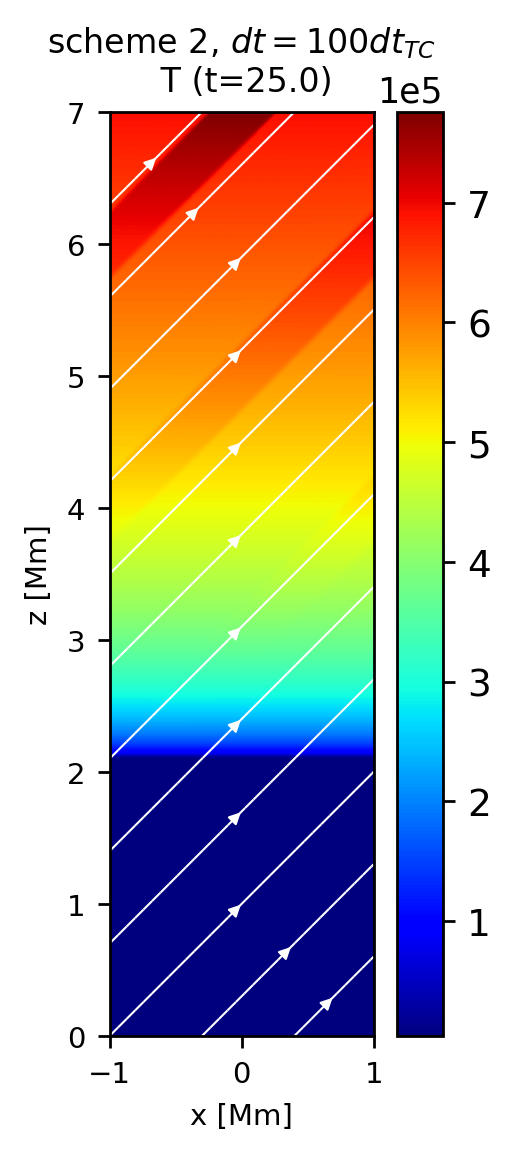}\\
\includegraphics[height = 0.21\textheight]{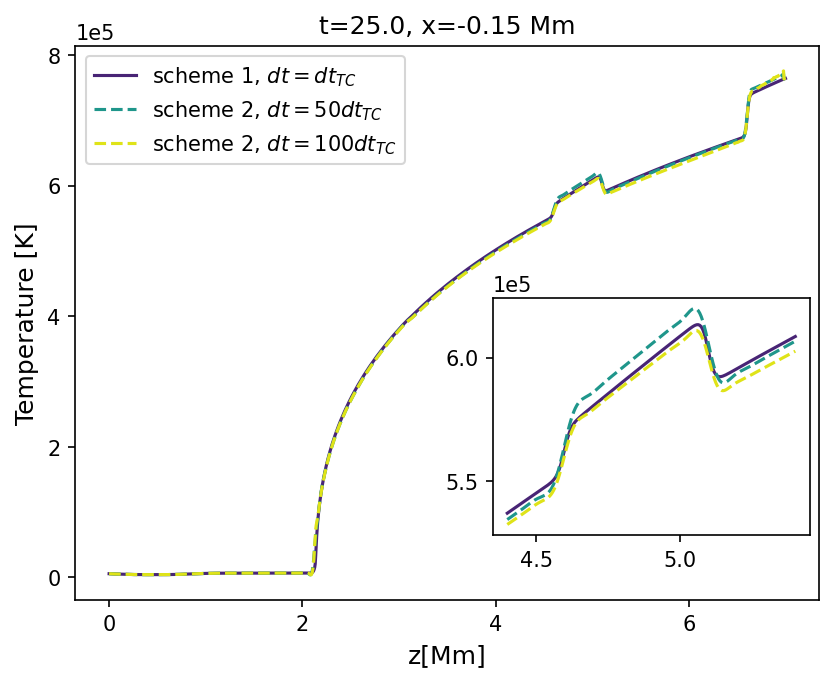}
\includegraphics[height = 0.21\textheight]{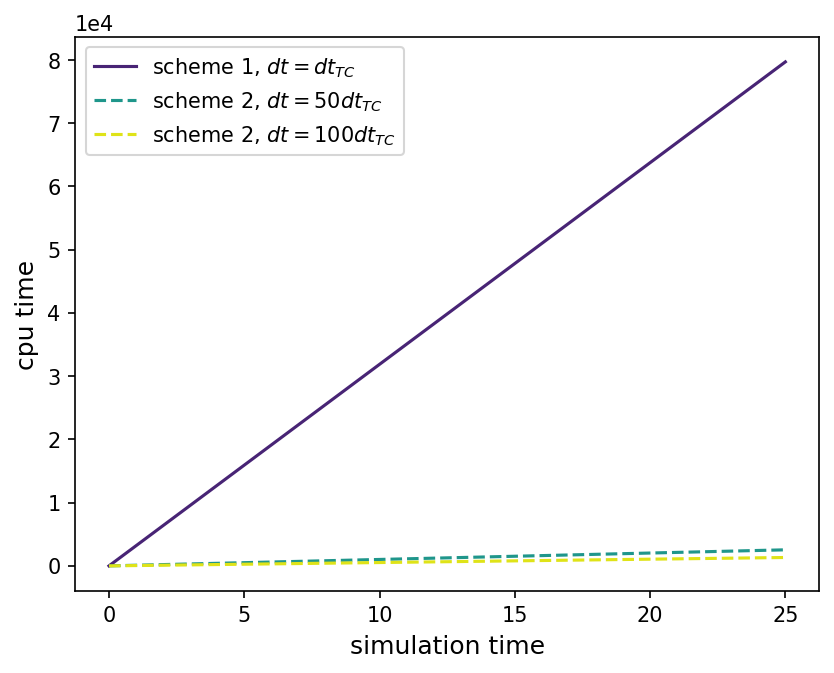}
\caption{Top panels: colormaps of the temperature and magnetic field lines at $t=25$ s obtained with the different schemes and time steps. Bottom left panel: 1D cut of the temperature. Bottom right panel: computational times. \label{fig:hottop}}
\end{figure*}

We have evolved this setup using the scheme \#1 with the time step $\mathrm{d}t = \mathrm{d}t_{\text{TC}}$, and with the scheme \#2 we set the time steps $\mathrm{d}t = 50 \mathrm{d}t_{\text{TC}}$ and $\mathrm{d}t = 100 \mathrm{d}t_{\text{TC}}$. The top panels of Figure \ref{fig:hottop} display colormaps of the temperature and magnetic field lines at $t = 25$ s obtained by the different schemes. It can be observed that, as expected the hot temperature is advected along the magnetic field lines: No important differences are observed between the results obtained by the different schemes. The bottom left panel shows a more detailed comparison of a 1D cut of the temperature at $x = -0.15$ Mm, where we find that the solutions exhibit a similar tendency.  Finally, the bottom right panel shows the cpu times for the different runs. The results suggest that the hyperbolic treatment can model heat conductivity under realistic coronal values, reducing significantly the computational cost. In these simulations the advection time step was 2893 times larger then the conduction one. For the simulation where the time step is $\mathrm{d}t = 50\mathrm{d}t_{\text{tc}}$, the efficiency ratio was of 31.38, and when $\mathrm{d}t = 100\mathrm{d}t_{\text{tc}}$ it was of 60.1. 

\section{Discussion and conclusions}
\label{sec:conclusions}

We have described the implementation of two numerical methods to model thermal conductivity in the code MANCHA3D. The scheme \#1 corresponds to the explicit evolution of the parabolic heat flux term, which leads to very restrictive time steps. Scheme \#2 is a hyperbolic treatment of the heat flux which allows the code to achieve considerably large speed-ups since it solves a separate hyperbolic equation for the parallel heat flux. Both schemes were tested in one, two and three spatial dimension by reproducing standard tests for anisotropic thermal conductivity.

We used the general heat flux expressions derived by \cite{1965RvPP....1..205B}. This formulation recovers the commonly used Spitzer expression in the limit of the strong magnetic field, but also takes into account the effects of the conductivity perpendicular and transverse to the magnetic field. The importance of the latter conductivity coefficients depends on the ratio between the collisional and cyclotron plasma frequencies, i.e. on plasma magnetization. 
This implementation is needed for modeling the whole solar atmosphere since it does a smooth transition between solar corona with remarkably anisotropic conductivity and the lower regions with large plasma $\beta$ and high collisionality where the heat flux becomes isotropic.

We propose to use a combination of both schemes, the parabolic treatment for the perpendicular and transverse conductivities and the hyperbolic treatment for the parallel conductivity, which is the one that imposes strict time steps due to its dependence on temperature. We show that in some regions of the atmosphere the perpendicular and transverse conductivities are not negligible. Namely, below the transition region, the perpendicular and transverse conductivities for ions and electrons might increase the heat flux. We present a two-dimensional test for the heat conduction in a stratified atmosphere, prove the robustness of the implemented methods to model thermal conduction, and confirm the significant speed up of the hyperbolic treatment over the computational cost of the simulations.

Our adaptation is similar to the one previously considered in the MURAM code \citep{2017ApJ...834...10R} and the PENCIL code \cite{2020GApFD.114..261W}. However, only the field aligned heat flux is considered in these two codes, which is, strictly speaking, valid only for the corona. In those works separate simulations from different zones of the atmosphere are coupled. 
The heat flux implemented in our code allows smooth transition from the corona to the convection zone, bringing us one step closer to realistic modeling of the solar atmosphere including the corona. 

A drawback of the adaptation we have used is that the hyperbolic scheme slows down the temperature evolution in regions of high temperature and low density like the corona, which would normally slow
down the code significantly, however this temperature variation should not affect much the physical results, as has been confirmed by \citep{2017ApJ...834...10R} and \cite{2020GApFD.114..261W}.  We also would like to point out that further analysis is still needed, since we have not considered complex magnetic configurations in an stratified atmosphere and we have not considered other non-ideal terms so far. In addition, an important issue that needs to be addressed in such complex scenarios is is how to increase correctly the magnitude of the timestep without reaching the unstable limit for the hyperbolic scheme. This matter will be addressed in our next project, in which we will study solar convection in a simulation box extended to the corona,
focusing on the interplay between the thermal conduction, the radiative losses and other non-ideal effects. We emphasize that the scheme will allow running simulations of the whole atmosphere with reasonable time steps, limited by advection processes below the chromosphere and not by the thermal conduction in the hottest part of the computational domain. Furthermore, we are planning to extend this work with the self-consistent two-fluid model developed by \cite{Peters_manuscript} which overcomes the shortcomings of the Braginskii formulation in the weakly collisional regime.

\section*{Acknowledgments}
 The authors thank the support by the European Research Council through the Consolidator Grant ERC-2017-CoG-771310-PI2FA and by the Spanish Ministry of Economy and the Industry and Competitiveness through the grant PGC2018-095832-B-I00. 

\bibliography{bibliography.bib}

\begin{thebibliography}{36}
\expandafter\ifx\csname natexlab\endcsname\relax\def\natexlab#1{#1}\fi

\bibitem[{Abdel-Hamid(1999)}]{ABDELHAMID1999899}
Abdel-Hamid, B. 1999, Applied Mathematical Modelling, 23, 899

\bibitem[{Balsara {et~al.}(2008)Balsara, Tilley, \&
  Howk}]{10.1111/j.1365-2966.2008.13085.x}
Balsara, D.~S., Tilley, D.~A., \& Howk, J.~C. 2008, Monthly Notices of the
  Royal Astronomical Society, 386, 627

\bibitem[{{Bingert} \& {Peter}(2011)}]{2011A&A...530A.112B}
{Bingert}, S. \& {Peter}, H. 2011, \aap, 530, A112

\bibitem[{{Bingert} \& {Peter}(2013)}]{2013A&A...550A..30B}
{Bingert}, S. \& {Peter}, H. 2013, \aap, 550, A30

\bibitem[{{Bourdin} {et~al.}(2013){Bourdin}, {Bingert}, \&
  {Peter}}]{2013A&A...555A.123B}
{Bourdin}, P.~A., {Bingert}, S., \& {Peter}, H. 2013, \aap, 555, A123

\bibitem[{{Bradshaw} {et~al.}(2012){Bradshaw}, {Klimchuk}, \&
  {Reep}}]{2012ApJ...758...53B}
{Bradshaw}, S.~J., {Klimchuk}, J.~A., \& {Reep}, J.~W. 2012, \apj, 758, 53

\bibitem[{{Braginskii}(1965)}]{1965RvPP....1..205B}
{Braginskii}, S.~I. 1965, Reviews of Plasma Physics, 1, 205

\bibitem[{Cattaneo(1958)}]{cattaneo1958forme}
Cattaneo, C. 1958, CR. Acad

\bibitem[{{Chen} {et~al.}(2014){Chen}, {Peter}, {Bingert}, \&
  {Cheung}}]{2014A&A...564A..12C}
{Chen}, F., {Peter}, H., {Bingert}, S., \& {Cheung}, M.~C.~M. 2014, \aap, 564,
  A12

\bibitem[{{Felipe} {et~al.}(2010){Felipe}, {Khomenko}, \&
  {Collados}}]{2010ApJ...719..357F}
{Felipe}, T., {Khomenko}, E., \& {Collados}, M. 2010, \apj, 719, 357

\bibitem[{{Fisher} {et~al.}(1985){Fisher}, {Canfield}, \&
  {McClymont}}]{1985ApJ...289..414F}
{Fisher}, G.~H., {Canfield}, R.~C., \& {McClymont}, A.~N. 1985, \apj, 289, 414

\bibitem[{{Gonz{\'a}lez-Morales} {et~al.}(2018){Gonz{\'a}lez-Morales},
  {Khomenko}, {Downes}, \& {de Vicente}}]{2018A&A...615A..67G}
{Gonz{\'a}lez-Morales}, P.~A., {Khomenko}, E., {Downes}, T.~P., \& {de
  Vicente}, A. 2018, \aap, 615, A67

\bibitem[{{Gudiksen} {et~al.}(2011){Gudiksen}, {Carlsson}, {Hansteen}, {Hayek},
  {Leenaarts}, \& {Mart{\'\i}nez-Sykora}}]{2011A&A...531A.154G}
{Gudiksen}, B.~V., {Carlsson}, M., {Hansteen}, V.~H., {et~al.} 2011, \aap, 531,
  A154

\bibitem[{{G{\"u}nter} {et~al.}(2005){G{\"u}nter}, {Yu}, {Kr{\"u}ger}, \&
  {Lackner}}]{2005JCoPh.209..354G}
{G{\"u}nter}, S., {Yu}, Q., {Kr{\"u}ger}, J., \& {Lackner}, K. 2005, Journal of
  Computational Physics, 209, 354

\bibitem[{Hunana {et~al.}(2022)Hunana, Passot, Khomenko, D., Collados,
  Tenerani, P., Maneva, Goldstein, \& Webb}]{Peters_manuscript}
Hunana, P., Passot, T., Khomenko, E., {et~al.} 2022, Manuscript in preparation

\bibitem[{Jiang {et~al.}(2012)Jiang, Fang, \& Chen}]{Jiang_2012}
Jiang, R.-L., Fang, C., \& Chen, P.-F. 2012, The Astrophysical Journal, 751,
  152

\bibitem[{{Khomenko} \& {Collados}(2006)}]{2006ApJ...653..739K}
{Khomenko}, E. \& {Collados}, M. 2006, \apj, 653, 739

\bibitem[{{Khomenko} \& {Collados}(2008)}]{2008ApJ...689.1379K}
{Khomenko}, E. \& {Collados}, M. 2008, \apj, 689, 1379

\bibitem[{{Khomenko} {et~al.}(2018){Khomenko}, {Vitas}, {Collados}, \& {de
  Vicente}}]{2018A&A...618A..87K}
{Khomenko}, E., {Vitas}, N., {Collados}, M., \& {de Vicente}, A. 2018, \aap,
  618, A87

\bibitem[{Liu {et~al.}(2009)Liu, Petrosian, \& Mariska}]{Liu_2009}
Liu, W., Petrosian, V., \& Mariska, J.~T. 2009, The Astrophysical Journal, 702,
  1553

\bibitem[{{Liu} {et~al.}(2022){Liu}, {Yuan}, {Xu}, {An}, {Zhao}, {Li}, \&
  {Yi}}]{2022JPCS..16110390L}
{Liu}, X., {Yuan}, S., {Xu}, B., {et~al.} 2022, Journal of Physics and
  Chemistry of Solids, 161, 110390

\bibitem[{{Meier} {et~al.}(2010){Meier}, {Lukin}, \&
  {Shumlak}}]{2010CoPhC.181..837M}
{Meier}, E.~T., {Lukin}, V.~S., \& {Shumlak}, U. 2010, Computer Physics
  Communications, 181, 837

\bibitem[{Meyer {et~al.}(2012)Meyer, Balsara, \&
  Aslam}]{10.1111/j.1365-2966.2012.20744.x}
Meyer, C.~D., Balsara, D.~S., \& Aslam, T.~D. 2012, Monthly Notices of the
  Royal Astronomical Society, 422, 2102

\bibitem[{{Navarro} {et~al.}(2017){Navarro}, {Lora-Clavijo}, \&
  {Gonz{\'a}lez}}]{2017ApJ...844...57N}
{Navarro}, A., {Lora-Clavijo}, F.~D., \& {Gonz{\'a}lez}, G.~A. 2017, \apj, 844,
  57

\bibitem[{{Navarro} {et~al.}(2021){Navarro}, {Lora-Clavijo}, {Murawski}, \&
  {Poedts}}]{2021MNRAS.500.3329N}
{Navarro}, A., {Lora-Clavijo}, F.~D., {Murawski}, K., \& {Poedts}, S. 2021,
  \mnras, 500, 3329

\bibitem[{{Parrish} \& {Stone}(2005)}]{2005ApJ...633..334P}
{Parrish}, I.~J. \& {Stone}, J.~M. 2005, \apj, 633, 334

\bibitem[{{Rempel}(2017)}]{2017ApJ...834...10R}
{Rempel}, M. 2017, \apj, 834, 10

\bibitem[{Sharma \& Hammett(2007)}]{SHARMA2007123}
Sharma, P. \& Hammett, G.~W. 2007, Journal of Computational Physics, 227, 123

\bibitem[{{Sharma} \& {Hammett}(2011)}]{2011JCoPh.230.4899S}
{Sharma}, P. \& {Hammett}, G.~W. 2011, Journal of Computational Physics, 230,
  4899

\bibitem[{{Spitzer}(1956)}]{Spitzer1956}
{Spitzer}, L. 1956, {Physics of Fully Ionized Gases} (Interscience Publishers)

\bibitem[{{Vernazza} {et~al.}(1981){Vernazza}, {Avrett}, \&
  {Loeser}}]{Vernazza_temperature_profile}
{Vernazza}, J.~E., {Avrett}, E.~H., \& {Loeser}, R. 1981, ApJS, 45, 635

\bibitem[{Vernotte(1958)}]{vernotte1958paradoxes}
Vernotte, P. 1958, Compt. Rendu, 246, 3154

\bibitem[{{Warnecke} \& {Bingert}(2020)}]{2020GApFD.114..261W}
{Warnecke}, J. \& {Bingert}, S. 2020, Geophysical and Astrophysical Fluid
  Dynamics, 114, 261

\bibitem[{{Xia} {et~al.}(2018){Xia}, {Teunissen}, {El Mellah}, {Chan{\'e}}, \&
  {Keppens}}]{2018ApJS..234...30X}
{Xia}, C., {Teunissen}, J., {El Mellah}, I., {Chan{\'e}}, E., \& {Keppens}, R.
  2018, \apjs, 234, 30

\bibitem[{{Ye} {et~al.}(2020){Ye}, {Shen}, {Lin}, \&
  {Mei}}]{2020A&C....3000341Y}
{Ye}, J., {Shen}, C., {Lin}, J., \& {Mei}, Z. 2020, Astronomy and Computing,
  30, 100341

\bibitem[{{Yu} {et~al.}(2022){Yu}, {Zhang}, {Zhao}, {Liu}, {Xia}, \&
  {Wu}}]{2022FrPhy..1723202Y}
{Yu}, H., {Zhang}, H., {Zhao}, J., {et~al.} 2022, Frontiers of Physics, 17,
  23202

\end{thebibliography}
\bibliographystyle{aa}

\end{document}